\documentclass[useAMS,usenatbib]{mn2e}

\usepackage{epsf,rotating}
\usepackage{amsmath}
\usepackage{subfigure}

\newcommand{\kmsmpc}{\kms\;{\rm Mpc}^{-1}}

\newcommand{\hkpc}{h^{-1}{\rm kpc}}
\newcommand{\hmpc}{h^{-1}{\rm Mpc}}

\newcommand{\kms}{\;{\rm km}\,{\rm s}^{-1}}

\newcommand{\cmc}{\;{\rm cm}^{-3}}

\newcommand{\gad}{{\sc Gadget-2}}

\newcommand{\ion}[2]{\hbox{#1\,{\sc #2}}}
\newcommand{\vw}{v_{\rm w}}

\newcommand{\tdep}{t_{\rm dep}}
\newcommand{\fgas}{f_{\rm gas}}

\title[Galaxy Evolution in Simulation II: Metals and Gas]{Galaxy Evolution in Cosmological Simulations with Outflows II: Metallicities and Gas Fractions}

\author[Dav\'e, Finlator, \& Oppenheimer]{
\parbox[t]{\textwidth}{\vspace{-1cm}
Romeel Dav\'e$^1$, Kristian Finlator$^2$, Benjamin D. Oppenheimer$^3$}
\\\\$^1$ Astronomy Department, University of Arizona, Tucson, AZ 85721, USA
\\$^2$ Hubble Fellow; Physics Department, University of California, Santa Barbara, CA 93106, USA
\\$^3$ Leiden Observatory, Leiden University, PO Box 9513, 2300 RA Leiden, Netherlands
}

\begin{document}

\maketitle

 \begin{abstract}
We use cosmological hydrodynamic simulations to investigate how
inflows, star formation, and outflows govern the the gaseous and
metal content of galaxies within a hierarchical structure formation
context.  In our simulations, galaxy metallicities are established
by a balance between inflows and outflows as governed by the mass
outflow rate, implying that the mass-metallicity relation reflects
how the outflow rate varies with stellar mass.  Gas content,
meanwhile, is set by a competition between inflow into and gas
consumption within the interstellar medium, the latter being governed
by the star formation law, while the former is impacted by both
wind recycling and preventive feedback.  Stochastic variations in
the inflow rate move galaxies off the equilibrium mass-metallicity
and mass-gas fraction relations in a manner correlated with star
formation rate, and the scatter is set by the timescale to
re-equilibrate.  The evolution of both relations from $z=3\rightarrow
0$ is slow, as individual galaxies tend to evolve mostly along the
relations.  Gas fractions at a given stellar mass slowly decrease
with time because the cosmic inflow rate diminishes faster than the
consumption rate, while metallicities slowly increase as infalling
gas becomes more enriched.  Observations from $z\sim 3\rightarrow
0$ are better matched by simulations employing momentum-driven wind
scalings rather than constant wind speeds, but all models predict
too low gas fractions at low masses and too high metallicities at
high masses.  All our models reproduce observed second-parameter
trends of the mass-metallicity relation with star formation rate
and environment, indicating that these are a consequence of equilibrium
and not feedback.  Overall, the analytical framework of our equilibrium
scenario broadly captures the relevant physics establishing the
galaxy gas and metal content in simulations, which suggests that
the cycle of baryonic inflows and outflows centrally governs the
cosmic evolution of these properties in typical star-forming galaxies.
\end{abstract}

\section{Introduction} 

Galaxies' stellar, gas, and metal content determine the majority
of their observable properties across all wavelengths.  Hence
understanding the co-evolution of these basic constituents is at
the heart of developing a comprehensive theory for the formation
and evolution of galaxies.  Advancing observations have now
characterised these properties in galaxies back into the peak epoch
of cosmic star formation and beyond.  Such observations provide
stringent tests for galaxy formation models, and offer new opportunities
to constrain the physical processes that govern galaxy evolution.

Observations have revealed tight correlations between stars, gas,
and metals in galaxies.  One example is the relationship between
stellar mass ($M_*$) and star formation rate (SFR), called the main
sequence for star-forming galaxies~\citep{noe07}; it is slightly
sub-linear and evolves with redshift roughly independently of
mass~\citep{elb07,dad07,dav08}.  Another example is the correlation
between stellar mass and gas-phase metallicity, called the
mass-metallicity relation~\citep[MZR; e.g.][]{tre04}, which shows
a remarkably low scatter across a wide range in mass~\citep{lee06,zhao10}.
Moreover, departures from the mean relation are strongly correlated
with other galaxy properties~\citep{ell08,coo08,pee09,lar10,man10}.
The scatter in metallicity is tightest when correlated with stellar
mass~\citep{tre04}, suggesting that stellar mass is primarily
responsible for governing the metal content of galaxies.  The gas
content is more difficult to measure because all phases must be
accounted for (atomic, molecular, and ionised), but still shows
a fairly tight anti-correlation with stellar mass~\citep{cat10,pee10}.
These relations evolve with redshift, towards lower
metallicity~\citep{sav05,erb06,zah11} and higher gas
content~\citep{erb06b,tac10} at higher redshift.

These trends are qualitatively consistent with the canonical scenario
for galaxy formation in which galaxies start out with a gaseous
halo that cools onto a central disk, forms stars, and self-enriches
while consuming its gas~\citep[e.g.][]{ree77,whi78}.  However, more
detailed observations have shown that the gas consumption rates
would exhaust the gas supply quickly, and therefore continual
replenishment of gas appears to be required to sustain star
formation~\citep{tac10,gen10,pap11}.  Moreover, the metallicity
evolution is quite slow, straining self-enrichment models~(e.g.
the classical G-dwarf problem).  Hierarchical structure formation
generically predicts inflow and merging that spur galaxy growth.
But unfettered inflow grossly overproduces global star formation,
known as the overcooling problem~\citep[e.g.][]{dav01,bal01}.
Therefore feedback processes must strongly regulate galaxy growth.
Such feedback processes are expected to manifest themselves in the
evolution of the mass, metal, and gaseous content of galaxies.
Hence understanding the origin and evolution of scaling relations
between these constituents provides key insights into accretion and
feedback processes that govern galaxy growth.

Cosmological hydrodynamic simulations have advanced rapidly over
the past decade, to a point where they can plausibly match a wide
range of properties of galaxies and the intergalactic medium (IGM)
across cosmic time.  One recently-explored physical process that
greatly improves concordance with observations is strong and
ubiquitous galactic outflows.  These outflows are powered by
supernovae, stellar winds, and/or photons from young stars, i.e.
they result from the star formation process itself, leading to
self-regulated growth.  Qualitatively, observations indicate that
galaxy formation must be increasingly suppressed towards small
masses.  Outflows are now directly observed in most star-forming
galaxies at $z\ga 1$~\citep[e.g.][]{wei09,ste10}.  By incorporating
outflows as observed into simulations, it is possible to yield
galaxy populations that significantly more closely resemble those
observed.

Recently, it has been found that simulations employing outflow
scalings as expected for momentum-driven winds~\citep{mur05,zha10}
are among the most successful at matching a wide range of data on
galaxies~\citep[e.g.][]{dav06,fin08,opp10} and the
IGM~\citep[e.g.][]{opp06,opp08,opp09,opp09b}, although notable
discrepancies remain.  These scalings assume that the mass outflow
rate scales inversely with galaxy circular velocity, providing
increased suppression of star formation in smaller systems.  Such
outflows also have interesting unanticipated consequences.  For
instance, \citet{opp08} found that ejected wind material more often
than not returned into galaxies, and that this so-called wind
recycling accretion becomes stronger at higher masses and dominates
the global accretion onto galaxies at $z\la 1$~\citep{opp10}.
Furthermore, winds also have a ``preventive" feedback effect
particularly in smaller galaxies, by adding energy to surrounding
gas which curtails inflow into the interstellar medium of galaxies
relative to inflow into the halo~\citep{opp10,vdv11,fau11}.  The
high outflow rates and frequent re-accretion suggest a continual
cycling of baryons between galaxies and their surrounding IGM, and
that this cycle is responsible for governing the observable properties
of both.

In this series of two papers, we investigate the way in which inflows
and outflows within a hierarchical structure formation context govern
the main constituents of galaxies, namely stars, gas, and metals.
In \citet[hereafter Paper~I]{dav11} we focused on stellar masses and star
formation rates.  We argue that many of the trends seen in simulations can
be understood within the framework of galaxies living in a slowly-evolving
equilibrium between inflow, outflow, and star formation.  The inflow is at
early epochs supplied primarily from the (relatively) pristine IGM, while
at later times wind recycling brings back gas in a mass-dependent fashion.
As in \citet{opp10}, we showed that outflows produce three-tiered
stellar mass and star formation rate functions, where the middle tier
is established by the onset of differential (i.e. mass-dependent)
wind recycling.  The evolution and mass dependence of the specific
star formation follows trends arising from the mass accretion rate into
halos, modulated by outflows.  We further examined the satellite galaxy
population, and found that in models they are not particularly more
common or more bursty than central galaxies at a given mass, and that
the main difference versus centrals is that satellites have increasingly
suppressed star formation to small masses.  We showed that momentum-driven
wind scalings provide the best overall fit to available observations,
but the agreement is only good in the range of $\sim (0.1-1)L^\star$.
At lower masses, star formation in dwarfs seems to occur at too early
epochs in the models, and at higher masses some additional mechanism is
required to quench star formation in massive galaxies~\citep[e.g. black
hole feedback;][]{dim05,cro06,del06,bow06,fon07,som08,gab11}.  Overall,
comparing these simulations to observations helps constrain the way in
which inflows and outflows work together to govern the growth of galaxies'
stellar component, while highlighting key failures of current models.

In this paper, Paper~II in this series, we extend the analysis of
our suite of cosmological hydrodynamic simulations with outflows
to examine galaxy metallicities and gas fractions.  The primary
goal is to understand how outflows govern scaling relations between
these quantities and their stellar content.  We will show that the
equilibrium scenario introduced in Paper~I also provides the basic
intuition for understanding gas and metal growth.  We outline a
simple analytic formalism that captures the main features of the
simulation results.  In it, the metallicity of galaxies is governed
primarily by outflows with a secondary effect from enriched infall,
while the gas content is governed by a competition between cosmological
gas supply and the gas consumption rate set by the star formation
law.  Both the metallicity and gas fraction are driven by cosmic
inflows, which diminish rapidly with cosmic time, and fluctuate on
shorter timescales resulting in deviations from the mean relations
that correlate with star formation.  By comparing to observations
we find that the momentum-driven wind scalings provide the best
match to data among the models examined here, but once again there
are significant discrepancies at the highest and lowest masses.
These results highlight how galactic outflows are a key moderator
of the stellar, metal, and gas content of galaxies at all epochs,
and in turn observations of these properties provide valuable
insights into the cosmic ecosystem within which galaxies form and
grow.

The paper is organized as follows. In \S\ref{sec:sims} we briefly describe
our hydrodynamic simulations including our galactic outflow models.
In \S\ref{sec:met} we examine simulated mass-metallicity relations across
cosmic time, and present a simple framework for understanding their
physical origin.  In \S\ref{sec:gas} we similarly examine galaxy gas
fractions.  In \S\ref{sec:datacomp} we compare to observations of metal
and gas content to identify broad constraints on feedback processes.
In \S\ref{sec:secondpar} we discuss second-parameter dependences of
the MZR and gas fractions with star formation rate and environment.
In \S\ref{sec:galevol} we explore how individual galaxies evolve in the
mass-metallicity and mass-gas fraction planes.  Finally, we summarize
and discuss the broader implications of our work in \S\ref{sec:summary}.

\section{Simulations}\label{sec:sims}

The suite of simulations employed are identical to that in Paper~I.
We refer the reader there for a full discussion of all details, and here
briefly review some of the key aspects.

\subsection{Runs}

Our simulations are run with an extended version of the \gad~N-body
+ Smoothed Particle Hydrodynamic (SPH) code \citep{spr05}.  We
assume a $\Lambda$CDM cosmology~\citep{hin09}: $\Omega_{\rm M}=0.28$,
$\Omega_{\rm \Lambda}=0.72$, $h\equiv H_0/(100 \kmsmpc)=0.7$, a
primordial power spectrum index $n=0.96$, an amplitude of the mass
fluctuations scaled to $\sigma_8=0.82$, and $\Omega_b=0.046$.  We
call this the r-series, where our general naming convention is
r[{\it boxsize}]n[{\it particles/side}][{\it wind model}].  Our
primary runs use a boxsize of $48\hmpc$ on a side with $384^3$ dark
matter and $384^3$ gas particles, and a softening length of
$\epsilon=2.5\hkpc$ (comoving, Plummer equivalent).  To expand our
dynamic range we run two additional sets of simulations with $2\times
256^3$ particles identical to the primary runs, except one having
a boxsize of $24\hmpc$ and $\epsilon=1.875\hkpc$, and the other
with a boxsize of $48\hmpc$ and $\epsilon=3.75\hkpc$.  SPH particle
masses are $3.6\times 10^7 M_\odot$, $1.5\times 10^7 M_\odot$, and
$12\times 10^7 M_\odot$ for the r48n384, r24n256, and r48n256 series,
respectively, and dark matter particles masses are approximately
$5\times$ larger.

Our version of \gad\ includes cooling processes using the primordial
abundances as described in \citet{kat96} and metal-line cooling as
described in \citet{opp06}.  We include heating from a metagalactic
photo-ionising flux from \citet{haa01}, but this has essentially no
effect on galaxies that we can resolve since their halo masses are
well above the mass where photo-ionisation strongly suppresses galaxy
formation~\citep[e.g.][]{oka08}.  Star formation is modeled using a
subgrid recipe introduced in \citet{spr03a} where a gas particle above a
density threshold of $n_{\rm H}=0.13 \cmc$ is modeled as a fraction of
cold clouds embedded in a warm ionized medium following \citet{mck77}.
Star formation (SF) follows a Schmidt law \citep{sch59} with the SF
timescale scaled to match the $z=0$ Kennicutt law \citep{ken98}.  We use
a \citet{cha03} initial mass function (IMF) throughout.  We account for
metal enrichment from Type II supernovae (SNe), Type Ia SNe, and AGB
stars, and we track four elements (C,O,Si,Fe) individually, as described
in \citet{opp08}.

Galactic outflows are implemented using a Monte Carlo approach
analogous to star formation.  Outflows are directly tied to the SFR,
using the relation $\dot M_{\rm wind}= \eta \dot M_{\rm SF}$, where
$\eta$ is defined as the mass loading factor.  The probabilities for a gas
particle to spawn a star particle are calculated from the subgrid model
described above, and the probability to be launched in a wind is $\eta$
times that.  If the particle is selected to be launched, it is given an
additional velocity of $v_w$ in the direction of {\bf v}$\times${\bf a},
where {\bf v} and {\bf a} are the particle's instantaneous velocity
and acceleration, respectively.  Choices of the parameters $\eta$
and $v_w$ define the ``wind model".  Once a gas particle is launched,
its hydrodynamic (not gravitational) forces are turned off until either
$1.95\times10^{10}/(\vw (\kms))$ years have passed or, more often, the gas
particle has reached 10\% of the SF critical density.  This attempts to
mock up chimneys generated by outflows that allow relatively unfettered
escape from the galactic ISM, and which are not properly captured by the
spherically-averaging SPH algorithm at $\ga$kpc resolution; it also yields
results that are less sensitive to numerical resolution~\citep{spr03b}.
For a further discussion of hydrodynamic decoupling, see \citet{dal08}
and Paper~I.

For this paper we run four wind models:\\
(i) {\bf No winds (nw)}, where we do not include outflows (i.e. $\eta=0$);\\
(ii) {\bf Constant winds (cw),} where $\eta=2$ and $\vw=680 \kms$ for all galaxies;\\
(iii) {\bf Slow winds (sw),} where $\eta=2$ and $\vw=340 \kms$ for all galaxies; and\\
(iv) {\bf Momentum-conserving winds (vzw),} where galaxies are identified
on-the-fly and their velocity dispersion $\sigma$ is estimated~\citep[see][]{opp08}, and then 
\begin{eqnarray}
  \vw &=& 3\sigma \sqrt{f_L-1}, \label{eqn: windspeed} \\
  \eta &=& \frac{\sigma_0}{\sigma} \label{eqn: massload},
\end{eqnarray} 
where $f_L=[1.05,2]$ is the luminosity factor in units of the
galactic Eddington luminosity (i.e. the critical luminosity necessary
to expel gas from the galaxy potential), and $\sigma_0=150$~km/s
is the normalization of the mass loading factor.  Choices for the
former are taken from observations~\citep{rup05}, while the latter
is broadly is constrained to match high-redshift IGM
enrichment~\citep{opp08}.  The velocity dispersion $\sigma$ is
estimated from the baryonic galaxy mass $M_{\rm gal}$ using~\citep{opp08}
\begin{equation}\label{eqn:sigma}
\sigma = 200 \Bigl(\frac{M_{\rm gal}}{5\times 10^{12} h^{-1} M_\odot} 
\frac{\Omega_m}{\Omega_b} \frac{H(z)}{H_0} \Bigr)^{1/3} \;\; {\rm km/s}.
\end{equation}
See Paper~I for further details.  We particularly note \S2.2 for a
discussion about issues related to the momentum budget in the vzw model.
To reiterate, while the energy budget of this model is well within that
provided by supernovae, the momentum required to eject gas in the vzw
model significantly exceeds the photon momentum input from young stars.
A physically realistic model would then need to either invoke gas that
is optically thick in the far infrared in order to extract additional momentum 
from photons after reprocessing by dust~\citep[e.g. as in the models of][]{hop11},
or else that it is a combination of supernovae and young stars that
drives the outflow~\citep[e.g. as forwarded by][]{mur10}.

\subsection{Computing galaxy metallicities and gas fractions}

We use SKID\footnote{http://www-hpcc.astro.washington.edu/tools/skid.html}
(Spline Kernel Interpolative Denmax) to identify galaxies as bound
groups of star-forming gas and stars~\citep{ker05,opp10}.  Our
galaxy stellar mass limit is set to be $\ge 64$ star
particles~\citep{fin06}, resulting in a minimum resolved mass of
$M_{\rm gal}=1.1\times 10^9 M_\odot$ in our r48n384 series of runs.
We will only consider galaxies with stellar mass $M_*\geq M_{\rm
gal}$ in our analysis.  We separate galaxies into central and
satellite galaxies by associating each galaxy with a halo, where
we identify halos via a spherical overdensity algorithm~\citep{ker05}.

To compute gas fractions, we set $M_{\rm gas}$ to be the mass of
all star-forming gas in a SKID-identified galaxy.  We then define
\begin{equation}
\fgas\equiv \frac{M_{\rm gas}}{M_{\rm gas}+M_*},
\end{equation}
where $M_*$ is the stellar mass of the galaxy.  Note that some
authors choose $\fgas=M_{\rm gas}/M_*$, and for instance,
\citet{pee10} argue that this definition is more natural in terms
of understanding the origin of the MZR.  Nevertheless, we prefer
the above definition because it intuitively translates into the fraction
of a galaxy's (baryonic) mass that is in gas.  In the end, so long as
models and data are compared using the same quantity, the exact
definition is not critical.

Our gas mass includes all phases of the ISM, including the cold neutral
medium, molecular gas, and the warm ionized medium.  The latter typically
makes a small mass contribution, but the relation between the first two
depends on the internal physics of the ISM (particularly self-shielding)
that our simulations do not accurately track~\citep[see][for further
discussion]{pop09}.  It is therefore important to compare to data that
accounts for both neutral (\ion{H}{i}) and molecular ($H_2$) components,
as well as having been corrected for Helium.  Furthermore, gas mass
measurements can be sensitive to surface brightness effects in the outer
regions of galaxies.

The determination of gas content in simulated galaxies depends on the
density threshold for star formation.  Here we assume this density
threshold is $n_H=0.13$~cm$^{-3}$, which is a somewhat arbitrary choice
motivated by resolution considerations~\citep{spr03b}.  This assumption
makes little difference for galaxy star formation histories, because
dynamical perturbations are typically so frequent that gas is rapidly
driven inwards until it forms stars, and thus raising (lowering) the
threshold would only introduce a small delay (advancement) in processing
gas into stars~\citep{sch10}.  Naively, gas fractions would be more
affected since changing the threshold density will include more or less
gas as ``star-forming".  But it is not straightforward to even determine
the sign of the effect.  For instance, lowering the threshold would
provide more gas eligible for star formation, but would also increase the
star formation rate, thus lowering the gas fraction.  In high-resolution
simulations of individual galaxies~\citep[e.g.][]{gue11}, it is seen that
raising the threshold density does lower the gas fraction, but this is
also sensitively dependent on their feedback algorithm which is quite
different than ours.  Hence gas fraction comparisons should be taken
with some caution, and are mainly intended to illustrate trends.

The galaxy gas-phase metallicity is defined as the SF-weighted metallicity
of all gas particles in SKID-identified galaxies.  This definition most
closely mimics how metallicities are measured observationally using
nebular emission lines emanating from star-forming regions.  We use the
oxygen abundance as a metallicity tracer in our models, since this is
also typical of observational determinations.  We assume a solar oxygen
mass fraction of 0.00574~\citep{asp09}, or [O/H]$_\odot+12=8.69$.
The weighting of metallicity by star formation mitigates the issues
regarding the outer regions of galaxies that plague gas fractions,
since star formation is typically concentrated in the central region of
the galaxy.


\section{Galaxy Metal Content}\label{sec:met}

In this section we will examine the drivers behind the mass-metallicity
relation (MZR) and its evolution out to high redshifts.  We will
focus on the physical mechanisms that connect inflows and outflows
to the observable metal content of galaxies, in particular placing
the form and evolution of the MZR within the context of the equilibrium
scenario for galaxy evolution.

\subsection{The $z=0$ mass-metallicity relation}\label{sec:massmet}

\begin{figure*}
\vskip -1.0in
\setlength{\epsfxsize}{0.85\textwidth}
\centerline{\epsfbox{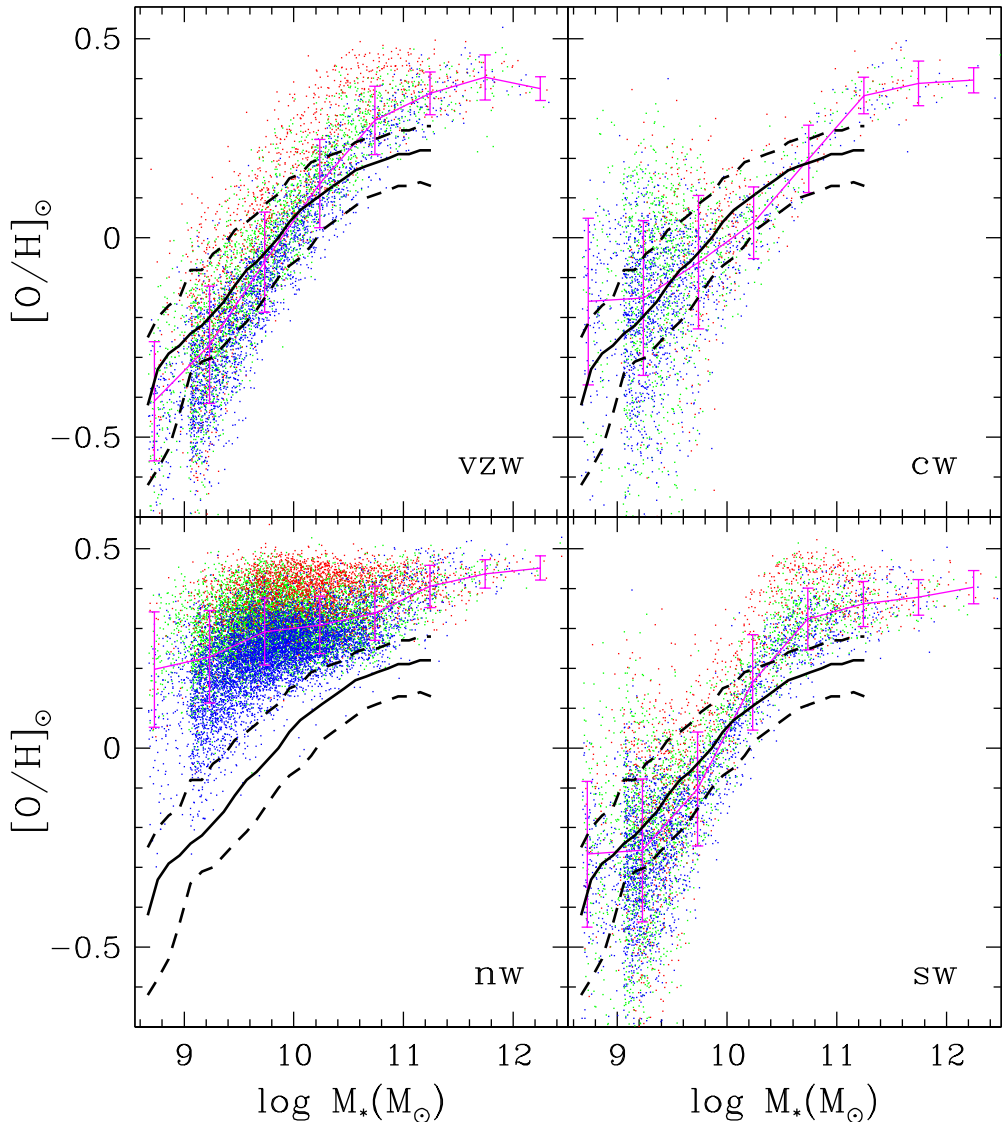}}
\vskip -1.0in
\caption{The $M_*-Z_{\rm gas}$ relation (MZR) at $z=0$ in our r48n384
cosmological hydrodynamic simulations employing our four galactic
outflow scalings: momentum-driven scalings (upper left), constant winds
(upper right), no winds (lower left), and slow winds (lower right).
Coloured points represent individual simulated galaxies, colour-coded by
SFR within bins of $M_*$ into upper (blue), middle (green), and lower
(red) thirds.  Magenta lines show a running median of the simulated
points, with $1\sigma$ scatter about the median.  The thick solid line
is the $z\approx 0$ MZR from SDSS~\citep{tre04} with dashed lines showing
the range enclosing $16\%-84\%$ of the data.  Note that all model oxygen
abundances have been multiplied by 0.8 in order to match the amplitude
of the observed MZR at $M_*\approx 10^{10}M_\odot$, which is within
systematic uncertainties in metallicity measures; the shape, scatter,
and evolution are independent predictions.
}
\label{fig:massmet}
\end{figure*} 

Figure~\ref{fig:massmet} shows the $z=0$ relation between stellar mass
and gas-phase metallicity, the MZR, in our simulations.  The simulation
data points are color-coded by star formation rate within bins of stellar
mass, which we will discuss in \S\ref{sec:secondpar}.  The main body
of points comes from the r48n384 runs; the smaller-volume r24n256 run
galaxies are shown as the sparser set of points at $M_*<1.1\times 10^9
M_\odot$ to extend the dynamic range.  The magenta curve shows a running
median, and the errorbars show $1\sigma$ deviations about the median.
For comparison, the SDSS mass-metallicity relation mean (thick line)
and $1\sigma$ scatter (dashed lines) is overlaid, but we will defer
discussion of comparisons to observations until \S\ref{sec:datacomp}.

Metallicity measures have an uncertain normalization.  This comes
from uncertainties in metallicity determinations~\citep[e.g.][]{kew08},
uncertainties in adopted metal yields~\citep[see discussion
in][]{opp08}, and uncertainties in the solar metal abundance~\citep{asp09}.
Hence we treat the overall metallicity normalization as a free
parameter.  Given that we do not have any form of feedback that
quenches massive galaxies in these runs~\citep[e.g.][]{gab11}, our
simulations most robustly model star-forming galaxies at masses
below $M^\star$.  Therefore we normalize our metallicities to the
observed MZR at $M_*\sim 10^{10}M_\odot$, where it so happens that all
our wind simulations predict roughly similar metallicities.  This
normalization requires us to multiply all simulated metallicities
by an arbitrary factor of 0.8.  We apply this same factor at all
redshifts and for all models.  Hence the independent predictions
of our simulations are the shape, slope, scatter, and evolution of
the MZR, but not its overall amplitude.

All the wind models produce a general trend of increasing metallicity
with mass and a turnover to flat at high masses.  The no-wind case
in contrast produces a nearly flat MZR, as one would expect from e.g.
closed box evolution.  All the wind models approach the no-wind case at
$M_*\ga 10^{11} M_\odot$, as these wind models all eject proportionally
less material from the most massive galaxies, and the material that is
ejected tends to quickly fall back in~\citep{opp10}.  If some
form of ejective feedback to quench massive galaxies was included in
our models~\citep[e.g.][]{gab11}, this plateau metallicity may be lower.

While broadly similar, there are clear differences between various
wind models.  The constant wind model yields another turnover at
low masses ($M_*\la 10^{10}M_\odot$) towards a flat MZR. The slow
wind model produces a similar turnover at somewhat smaller masses.
The momentum-driven scalings case does not produce such a flattening,
at least within the mass range probed by these simulations; the MZR
slope here is nearly constant at $Z\propto M_*^{0.3}$ for $M_*\la
10^{10.5}M_\odot$.

To understand the origin of these features for the various wind
models, we review the findings from \citet{fin08} who developed a
simple analytic understanding of the MZR.  In their model, the
gas-phase metallicity of a galaxy is set by a balance between inflow
and outflow plus star formation.  Inflow brings in low-metallicity
gas, star formation enriches that gas, while outflows modulate the
fraction of inflow that turns into stars.  In equilibrium, the three
terms are related by
\begin{equation}\label{eqn:equil}
\dot{M}_{\rm inflow}=\dot{M}_{\rm *}+\dot{M}_{\rm outflow}
\end{equation}
\citep[see also][]{dut10,bou10}.
Rewriting this in terms of the mass loading factor
$\eta\equiv\dot{M}_{\rm outflow}/\dot{M}_{\rm *}$, we obtain
\begin{equation}\label{eqn:min}
\dot{M}_{\rm inflow}= (1+\eta)\dot{M}_{\rm *}
\end{equation}
In this simple ``equilibrium" picture, the metallicity is given by
the amount of metals produced by star formation, which is $y\dot{M}_*$
where $y$ is the yield of metals per unit star formation, divided by
the rate of gas inflow to be enriched.  Hence
\begin{equation}\label{eqn:mzr}
Z = y\dot{M}_*/\dot{M}_{\rm inflow} = \frac{y}{1+\eta},
\end{equation}
Here we have, for simplicity, assumed that the infalling gas has negligible
metallicity; we will relax this assumption in \S\ref{sec:galevol} 
(see Equation~\ref{eqn:mzralpha}).

A key assumption in this formalism is that the mass loading factor
reflects the amount of material that is ejected from the galaxy
without returning quickly.  In this sense, $\eta$ should be regarded as an
``effective" mass loading factor, which \citet{fin08} showed generally
tracks the input value of $\eta$ in our simulations when the wind velocity
is comparable to or exceeding the escape velocity.  We reiterate that
our simulations hydrodynamically decouple outflowing gas; simulations
that choose not do so can have significantly different effective
$\eta$ despite having the same input $\eta$~\citep[e.g.][]{dal08}.
The similarity between our input $\eta$ and effective $\eta$ (at least
above the escape velocity) thus reflects this particular modeling choice.

It is worth noting that Equation~\ref{eqn:mzr} does not have any explicit
dependence on stellar mass, but only depends on inflow and outflow
rates.  The physical interpretation is that the gas-phase metallicity
does not reflect a historical record of star formation in a galaxy (as
in a closed-box scenario), but rather reflects its recent (i.e. over
a gas depletion timescale; \S\ref{sec:tdep}) balance between inflows
and outflows.  This then distinguishes stellar metallicities, which must
reflect the history of metal buildup, from gas-phase ones.  In practice,
however, the fairly slow evolution of the MZR (\S\ref{sec:mzevol})
means that galaxy stellar metallicities are only slightly lower than
gas-phase ones.  We leave a detailed examination of stellar vs. gas-phase
metallicities for future work.

Using Equation~\ref{eqn:mzr}, we can understand the behavior of the
various wind models.  In the no-wind case, $\eta=0$, and the metallicity
is therefore close to constant.  Although the MZR is close to flat,
there remains a slightly slope owing to the rapidity of infall and the
lack of reduction of fresh gas by outflows, which results in galaxies
being not quite able to attain equilibrium.  We will see in the next
section that this effect becomes exacerbated at higher redshifts.
Furthermore, there is more enriched infall into higher mass galaxies,
as we discuss in \S\ref{sec:galevol}.

The momentum-driven wind scalings assume $\eta\propto v_c^{-1}\propto
M_*^{-1/3}$ (approximately).  Hence when $\eta\gg 1$, this approximates
$Z\propto M_*^{1/3}$.  The turnover at high masses is set by the
normalization of $\eta$, namely $\sigma_0$, which produces $\eta\la 1$
for $M_*\ga 10^{11} M_\odot$ (at $z=0$; this mass evolves mildly upwards
with redshift).  

The constant and slow wind cases introduce another consideration:
The competition between wind speed and escape velocity.  Unlike in the
momentum-driven scalings where $\vw\sim v_{\rm esc}$, here there is a
transition mass above which the winds cannot escape, and fall quickly back
into the galaxy.  Hence the wind recycling time is very short, meaning
that winds have little effect~\citep{opp10}.  Stated in terms of the above
formalism, the effective mass loading factor in these models approach zero
above that threshold mass (see \citealt{fin08} and \S2.2 of Paper~I), so
that at low masses $\eta\approx 2$ while at high masses $\eta\approx 0$,
with a steep transition between these regimes across which the metallicity
changes by a factor of $1+\eta=3$.  Because the wind speed is twice as
fast in the cw case, the transition occurs at a higher mass than in sw.
Although the transition to the low-mass regime is not well probed at
the dynamic range of these simulations (particularly in the sw case),
it is still evident.  This transition is also evident in the stellar
mass and star formation rate functions in Paper~I.  This equilibrium
scenario strongly predicts that the low-mass MZR will continue to be
flat to small masses in this model.

These simulations and the associated equilibrium model indicate that
the MZR is critically governed by the mass outflow rate from galaxies
(i.e. $\eta$) and its scaling with $M_*$.  This differs fundamentally
from the canonical explanation for the MZR that invokes a competition
between the galaxy potential well and the outflow velocity to modulate the
metals retained within a galaxy~\citep[e.g.][]{dek86,dek03,tre04}; here,
there is no mention of the potential well depth except indirectly via its
effects on the mass loading factor.  It is often canonically stated that
low mass galaxies preferentially eject more of their metals, and hence
have lower metallicity.  In our scenario, it is not the {\it ejection}
of metals that is modulated by outflows, it is the {\it production}
of metals from a given amount of hierarchical inflow that is regulated
by outflows.

The turnover in the MZR at high masses has typically been thought to
reflect the transition to a regime where outflows can no longer escape
the galaxy~\citep{dek86,tre04}.  In our constant-$\eta$ cases, this
is accurate.  But in our momentum-driven wind scalings case, since
the wind speed scales with the escape velocity.  Hence the turnover
is instead caused by $\eta$ becoming less than unity at large $M_*$ in
Equation~\ref{eqn:mzr}.  Since this model seems to predict an MZR that
is overall in better agreement with data, particularly to low masses,
this suggests that the conventional interpretation of the MZR being
governed by the potential well depth may not be accurate.

In summary, the MZR is governed by the ``star formation efficiency",
where here we mean this in the cosmological sense as the amount of
infalling material that is converted into stars\footnote{This is
notably different than the definition of star formation efficiency
in the interstellar medium, which describes how quickly molecular
gas is processed into stars.}.  This interpretation agrees with the
simulations of \citet{bro07}, who also found that the MZR is governed
by the cosmological star formation efficiency that is modulated by
supernova feedback.  In our models, this star formation efficiency and
hence the metallicity is directly controlled by the rate at which mass
is ejected in outflows.

\subsection{MZR evolution}\label{sec:mzevol}

\begin{figure*}
\vskip -0.1in
\setlength{\epsfxsize}{0.85\textwidth}
\centerline{\epsfbox{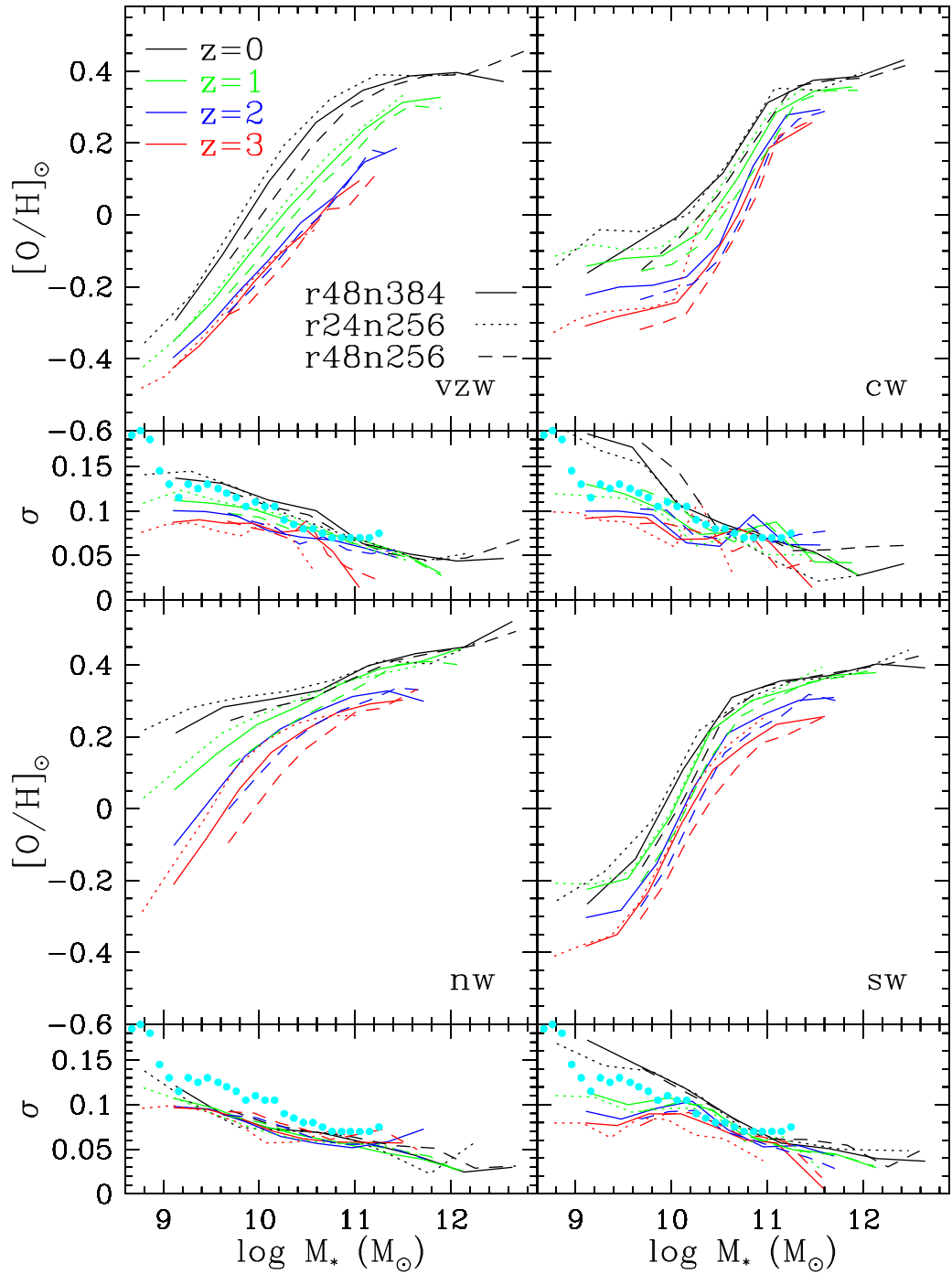}}
\vskip -0.5in
\caption{Large panels show the evolution of the mass-metallicity relation
in our four wind models at $z=0,1,2,3$.  Lines show the running median
within mass bins at each redshift.  Solid lines show the results from our
r48n384 simulations, dotted lines show r24n256 runs, and dashed lines
show r48n256 runs; the general consistency between the three indicates
that the results are numerically converged, though the lowest resolution
runs show noticeably lower metallicities.  All models show an upwards evolution
of metallicity at a given mass (at a rate that depends on wind model),  
although the characteristic MZR shape unique to each model does not change 
with redshift.  Smaller panels below each large panel show the $1\sigma$
scatter about the median relation for all the models.  Cyan points
show the $\approx 1\sigma$ scatter in observations of the $z=0$ MZR
from \citet{tre04}.
}
\label{fig:mzevol}
\end{figure*} 

Figure~\ref{fig:mzevol} depicts the evolution of the MZR from
$z=3\rightarrow 0$ in our four wind models.  We show the running median
at each redshift in the large panel, and the small panel below shows the
$1\sigma$ variance about the median within each mass bin.  The cyan points
in the lower panels show the observed $1\sigma$ variance at $z\approx
0$ from SDSS data~\citep{tre04} for comparison; we will discuss this
further in \S\ref{sec:secondpar}.  We separately show the results for
the r48n384 (solid lines), r24n256 (dotted), and r48n256 (dashed) runs;
the good agreement at all overlapping masses indicates that these results
are resolution-converged at least over the range of resolutions and
volumes probed here.  Although we don't show it here, all models at all
epochs retain the second-parameter trend shown in Figure~\ref{fig:massmet}
wherein lower SFR galaxies at a given $M_*$ have higher metallicities.

All the wind models have the general shape of their MZR as expected from
the equilibrium model and described in the previous section.  The shapes
remain similar at all redshifts, because the form of $\eta(M_*)$
for each model does not change, and this governs the MZR shape as
described in \S\ref{sec:massmet}.  The no-wind case appears further
out of equilibrium at earlier epochs, as it deviates more strongly from
the expected behavior of a constant metallicity at all masses.  This is
expected because accretion rates are higher at early epochs in comparison
to gas processing rates within the ISM; we will discuss this further in
\S\ref{sec:fgevol}.  The trend of a metallicity increasing with time at a
given $M_*$ is quite generic, at least among the models considered here.
This is not trivial; it is possible to design models that are quite
reasonable in many ways but yield the opposite evolution~\citep[][and
M. Arrigoni, priv.  comm.]{arr10}.

In detail, the evolutionary rate at a given $M_*$ varies somewhat
with wind model.  The momentum-driven wind scalings produce little
early evolution, and more evolution from $z\sim 2\rightarrow 0$.
The constant-$v_w$ models have less late evolution, particularly in the
slow wind case, and more rapid evolution at early epochs.  In Paper~I we
saw that such trends are also seen in the evolution of the galaxy star
formation rate functions (Figure~2 of Paper~I), where the constant-$v_w$
cases evolved less out to $z\sim 2$ compared to the momentum-driven
scalings case.  This qualitative similarity in evolution is consistent
with the interpretation that the MZR is governed primarily by galaxies'
star formation rates, as suggested by Equation~\ref{eqn:mzr}.  We will
discuss the evolution of the MZR further in \S\ref{sec:galevol}, when we
study how individual galaxies evolve in mass-metallicity space.

\section{Galaxy Gas Content}\label{sec:gas}

In this section we examine galaxy gas fractions and their evolution
across cosmic time.  As with the MZR, we attempt to provide physical
intuition for what establishes a galaxy's gas fraction and its
evolution by connecting it to gas inflow and outflow processes.

\subsection{Gas fractions at $z=0$} \label{sec:fgas}

\begin{figure*}
\vskip -1.0in
\setlength{\epsfxsize}{0.85\textwidth}
\centerline{\epsfbox{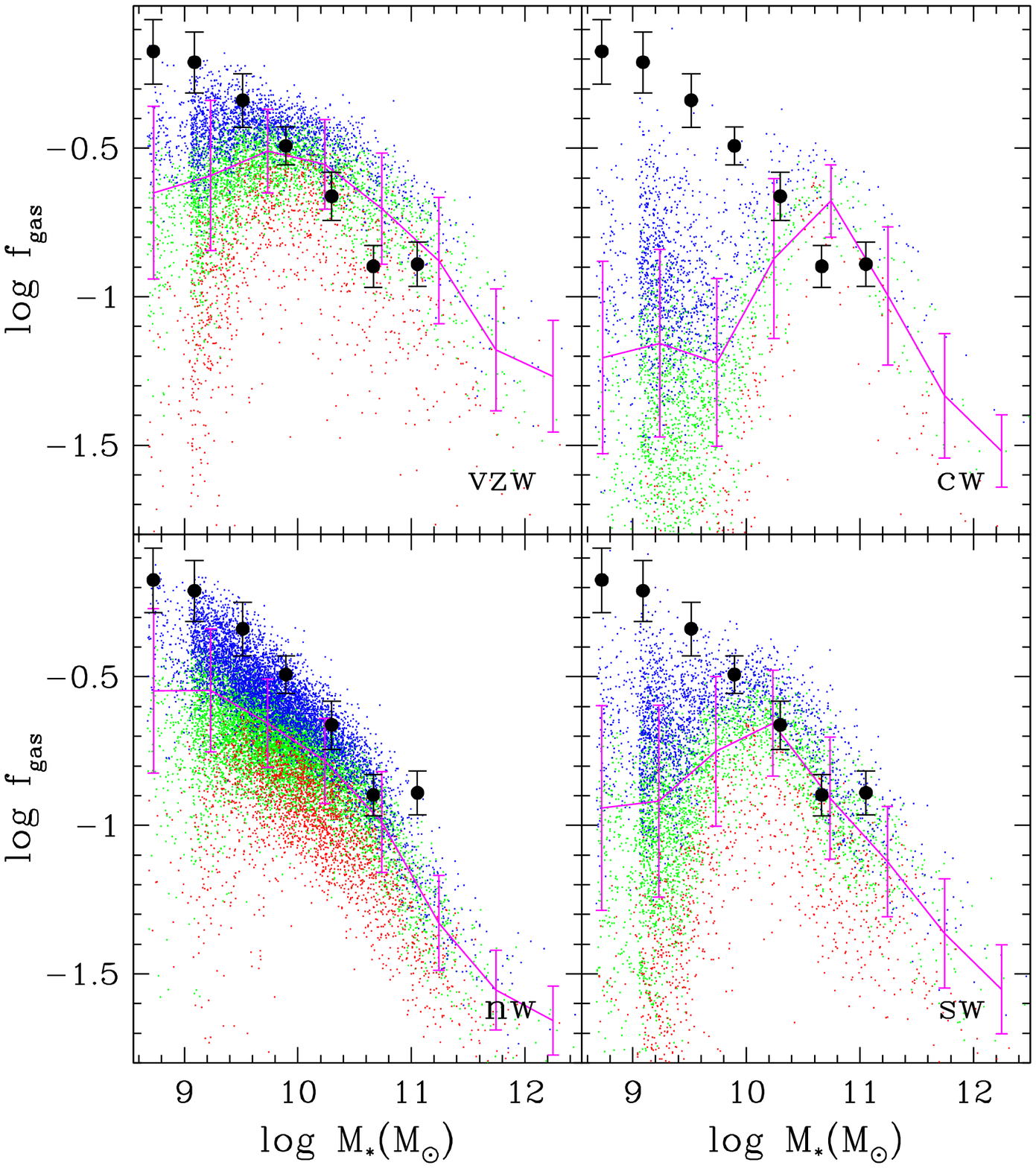}}
\vskip -1.0in
\caption{ The relation between $\fgas$ and $M_*$ (MGR) at
$z=0$ in our four winds models, with magenta lines showing the 
median and $1\sigma$ scatter as in Figure~\ref{fig:massmet}.
The points are color-coded by SFR within stellar mass bins: blue for
upper third, green for middle third, red for bottom third.
Data points show mean values as a function of $M_*$ from a
compilation of observations by \citet{pee10}.
}
\label{fig:fgas}
\end{figure*} 

Figure~\ref{fig:fgas} shows the $z=0$ gas fractions in our suite
of simulations as a function of stellar mass (the mass-gas relation,
or MGR).  The points are color-coded by star formation within a
given mass bin as in Figure~\ref{fig:massmet}; this will be discussed
in \S\ref{sec:secondpar}.  For comparison, the data points with
errors show observed gas fractions (\ion{H}{i}+H$_2$), corrected
for Helium, compiled and binned by \citet{pee10}.

At the massive end, all models show decreasing gas fractions with stellar
mass.  At lower masses, however, all wind models eventually deviate from
this trend, displaying a maximum typical gas fraction below which the
gas fraction becomes lower to smaller masses.  The no-wind case shows
no such maximum.  The mass at which this maximum occurs appears to be
related to wind recycling, i.e. the return of previously-ejected material
back into a galaxy.  \citet{opp10} showed that the recycling time becomes
long at smaller masses, eventually exceeding a Hubble time.  The mass
at which the recycling time becomes longer than the Hubble time in each
wind model is, to a good approximation, the mass at which the maximum
gas fraction is seen.  This suggests that the lower gas fractions at
small masses occurs at least in part because ejected winds never return
to these galaxies.

As emphasized by \citet{vdv11} and \citet{fau11}, feedback adds energy
to surrounding gas and prevents material from entering into smaller
galaxies.  Since much of the material entering into galaxies' ISM at
$z=0$ is recycled winds~\citep{opp10}, the majority of the effect is
that small galaxies do not re-accrete their winds.  The constant wind
model shows the highest turnover mass in $\fgas$, while the slow wind and
momentum-driven scalings cases occur at lower masses.  The latter wind
model also shows a somewhat slower drop-off in $\fgas$ to lower masses,
reflecting its less steep dependence of recycling time with mass.

Another possible issue is that perhaps many low-mass galaxies are
satellites that have lower gas content owing to stripping or strangulation
processes.  However, we will show in \S\ref{sec:sat} that the satellite
fraction does not increase appreciably to small masses, and we will
demonstrate that the turn-down in gas fractions to small masses is
present in both satellites and centrals.  Hence this particular trend
does not arise from satellites, although other interesting trends do
that we will explore in \S\ref{sec:sat}.

The non-monotonic behavior of gas fractions in wind models reflects a
similar behavior, with similar characteristic scales, as the specific
star formation rate (sSFR$\equiv$SFR$/M_*$) examined in Paper~I.
There we saw that low-mass galaxies had depressed sSFRs relative to an
extrapolation from higher masses in all wind models; the no-wind case
showed no such deviation.  This trend arises because of a combination of
wind recycling, which brings extra accretion at high masses~\citep{opp10},
and preventive feedback which suppresses accretion into galaxies at the
lowest masses~\citep[e.g.][]{vdv11}.  The suppression of gas fractions
is seen to be directly proportional to the suppression in sSFRs, which
we will explain in \S\ref{sec:fgevol}.  Hence the discrepancies of models
relative to observed sSFRs of dwarf galaxies noted in Paper~I is directly
traceable to lowered gas fractions in dwarfs predicted in models.

In summary, gas fractions fall with mass at the highest masses but
show a turnover at low masses in all the wind models.  This turnover is
not seen in observations, as we will discuss in \S\ref{sec:datacomp}.
To understand the origins of these trends, and also gas fraction
evolution, let us examine an instructive quantity for understanding gas
processing in galaxies, namely the depletion time.

\subsection{Depletion time}\label{sec:tdep}

\begin{figure*}
\vskip -0.5in
\setlength{\epsfxsize}{0.85\textwidth}
\centerline{\epsfbox{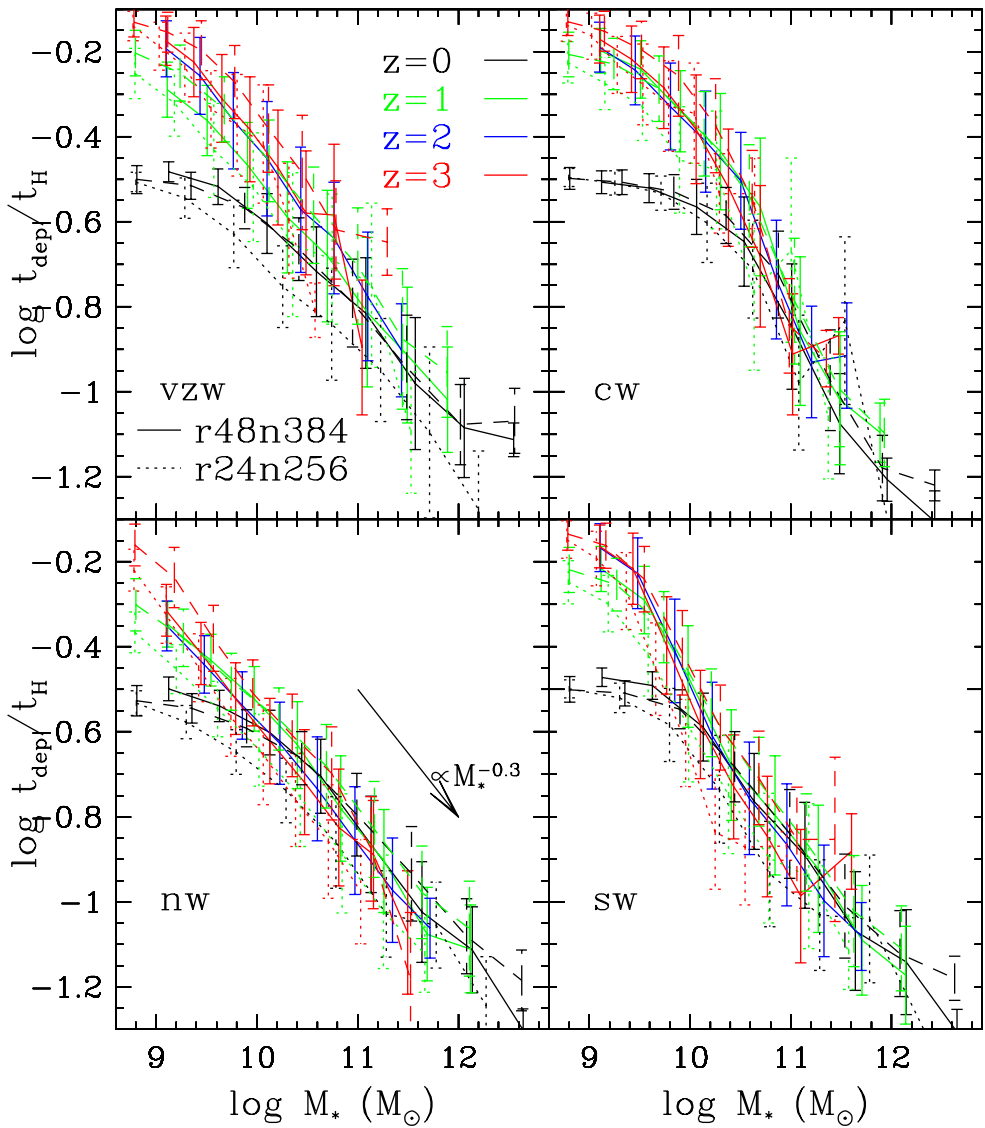}}
\vskip -0.5in
\caption{Depletion time $\tdep\equiv M_{\rm gas}$/SFR in units
of the Hubble time $t_H$ in our four wind models at $z=0,1,2,3$.  
Solid lines show results for the r48n384 series, dotted lines show
r24n256, and dashed lines show r48n256.  $\tdep$ 
drops to larger masses roughly as $M_*^{-0.3}$
and scales with $t_H$, essentially independent of winds, as derived
in Equation~\ref{eqn:tdep}.
}
\label{fig:tdep}
\end{figure*} 

Figure~\ref{fig:tdep} shows the gas depletion time $\tdep$ as a
function of stellar mass in our four wind models, from $z=3\rightarrow
0$.  We define $\tdep\equiv M_{\rm gas}/$SFR, which is the time
that galaxy would take to consume its current gas supply at its
current star formation rate.  Our three sets of simulations for
each wind model are shown to demonstrate good resolution convergence
over the dynamic range probed.  To illustrate a useful trend, we
divide $\tdep$ by the Hubble time $t_H$ at each redshift.

Several general features of $\tdep(M_*,z)$ are evident in
Figure~\ref{fig:tdep}:  First, higher-mass galaxies have shorter
depletion times.  Second, depletion times are, to first order,
independent of wind model.  Finally, $\tdep/t_H$ is essentially
invariant over most of cosmic time.  The insensitivity to wind
model and invariance when scaled to the Hubble time provide key
clues into the physics governing $\tdep$, and in turn, galaxy
gas contents.

We can broadly understand these trends using a straightforward argument
based on our star formation law.  The depletion time measures how gas,
once within a galaxy, gets consumed into stars.  Our simulations model
conversion of gas to stars by assuming a Kennicutt-Schmidt Law, which
equivalently follows the relation that the star formation rate is the gas
mass divided by the dynamical time $t_{\rm dyn}$ (of the star-forming
disk), times some overall efficiency factor that is measured to be
around 2\% both locally and in distant galaxies~\citep[e.g.][]{gen10}.
Hence the depletion time should scale as the dynamical time.  In a
canonical disk model~\citep{mo98}, the dynamical time evolves as $t_H$,
and hence we expect $t_{\rm dep}/t_H$ to be approximately constant,
which is generally confirmed in Figure~\ref{fig:tdep}, although with
deviations at low masses at $z=0$ that we discuss below.

\begin{figure}
\vskip -0.3in
\setlength{\epsfxsize}{0.65\textwidth}
\centerline{\epsfbox{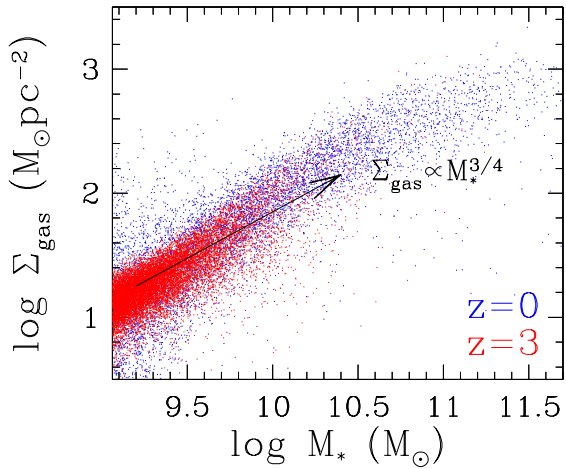}}
\vskip -3.0in
\caption{Gas surface density $\Sigma_{\rm gas}$ (comoving) versus stellar 
mass $M_*$ in galaxies from our r48n384vzw simulation, at $z=0$ (blue) and 
$z=3$ (red).  The gas surface density is taken to be the gas mass divided
by $\pi R^2$, where $R$ is the stellar half-mass radius of the galaxy.
This relation is reasonably well described by $\Sigma_{\rm gas}\propto
M_*^{3/4}$, as shown by the arrow.
}
\label{fig:sigmagas}
\end{figure} 

The mass dependence of $\tdep$ depends on the conversion rate of gas
into stars.  This is set by the details of internal structure of galaxies
within our simulation along with the assumed Kennicutt-Schmidt star
formation law, which is $\Sigma_{\rm SF}\propto \Sigma_{\rm gas}^{1.4}$
where $\Sigma_{\rm SF}$ and $\Sigma_{\rm gas}$ are the surface densities
of star formation and star-forming gas, respectively.  Hence $\tdep=
\Sigma_{\rm gas}/\Sigma_{\rm SFR}\propto \Sigma_{\rm gas}^{-0.4}$.  

We then employ an empirical relation measured in our
simulations of $\Sigma_{\rm gas}\propto M_*^{3/4}$ that we show in
Figure~\ref{fig:sigmagas}.  It does not evolve appreciably with redshift.
While we do not explicitly show it, this relation holds reasonably well
for all the wind models.  Observations of the stellar mass density
profile in late-type SDSS galaxies indicate $\Sigma_{\rm *}\propto
M_*^{0.54}$~\citep{kau03}, which is slightly shallower.  Gas profiles
are more difficult to measure, but at least in our models, star-forming
gas generally traces stars.  Using this shallower slope would not
significantly change our results, but we prefer to use the simulated
slope since we are trying to develop an analytic understanding of the
simulation results.  Our r24n256 and r48n256 simulations have slightly
different amplitudes for this relation but the slope is identical,
suggesting that the trend with $M_*$ is insensitive to resolution at
least over the (admittedly narrow) range probed by these simulations.

Putting this together, we obtain
\begin{equation}\label{eqn:tdep}
\tdep \propto t_H M_*^{-0.3}.
\end{equation}
This scaling provides a good match to $\tdep(M_*)$ in our simulations
at most epochs and masses as shown in Figure~\ref{fig:tdep}.  Since
we assume the same star formation law in all our wind models, there
is little sensitivity to outflows in this relation.

We note that $\tdep\propto t_H$ does not necessarily have to be the
case -- for instance, in mergers, $t_{\rm dyn}$ is quite small
within the dense central star-forming region during the starburst
phase, meaning that $\tdep$ of such starburst galaxies should lie
below the mean relation.  The relatively tight relation of $\tdep(M_*)$
in our models indicates that, in analogy with the tight main
sequence~\citep[Paper I]{dav08}, mergers are not a dominant
population at any redshift.  Hence even though we have not explicitly
used the equilibrium condition in deriving $\tdep$, the connection
between global virialization and the properties of the star-forming
region in galaxies only holds when galaxies are in a steady-state
situation.

$\tdep$ is to first order insensitive to outflows; all models, regardless
of winds, show a roughly similar $\tdep(M_*)$.  This is expected since
the derivation of Equation~\ref{eqn:tdep} does not involve any aspect that
depends on outflows, effectively employing only the star formation law and
virial arguments.  This supports the idea of Paper~I that star formation
is supply-regulated, that is, star formation occurs in proportion to the
gas available to form stars.  This differs relative to expectations from
models in which galaxies begin with a large reservoir of gas and consume
them rapidly~\citep[e.g.][]{egg62,mar10}.

At $z=0$, we see a systematic departure from the mean trend towards lower
$\tdep$ at low masses.  Part of this owes to satellite galaxies that are
increasingly starved of gas to low masses at low redshifts~(Figure 6 of
Paper~I), which occurs even in the no-wind case.  In the wind models,
an additional role is played by preventive feedback and the lack of wind
recycling at small $M_*$.  As usual for effects involving wind recycling,
the departure from the canonical relation occurs at a higher mass in cw
relative to sw, and is more gradual in the vzw case.  These effects are
very mild at earlier epochs when wind recycling is not as common.

The depletion time has sometimes been interpreted as a measure of
star formation efficiency variation with galaxy mass, such that
more massive galaxies more efficiently convert gas into stars.  But
in our models, the star formation efficiency is an input constant
that is calibrated to match observations of present-day
disks~\citep{spr03b,opp08}, and does not vary with galaxy mass.
The trend of $\tdep$ with $M_*$ arises from the assumed star formation
formation law, which is why it is very weakly dependent on details
of feedback.

\subsection{Gas fraction evolution}\label{sec:fgevol}

\begin{figure*}
\vskip -0.2in
\setlength{\epsfxsize}{0.85\textwidth}
\centerline{\epsfbox{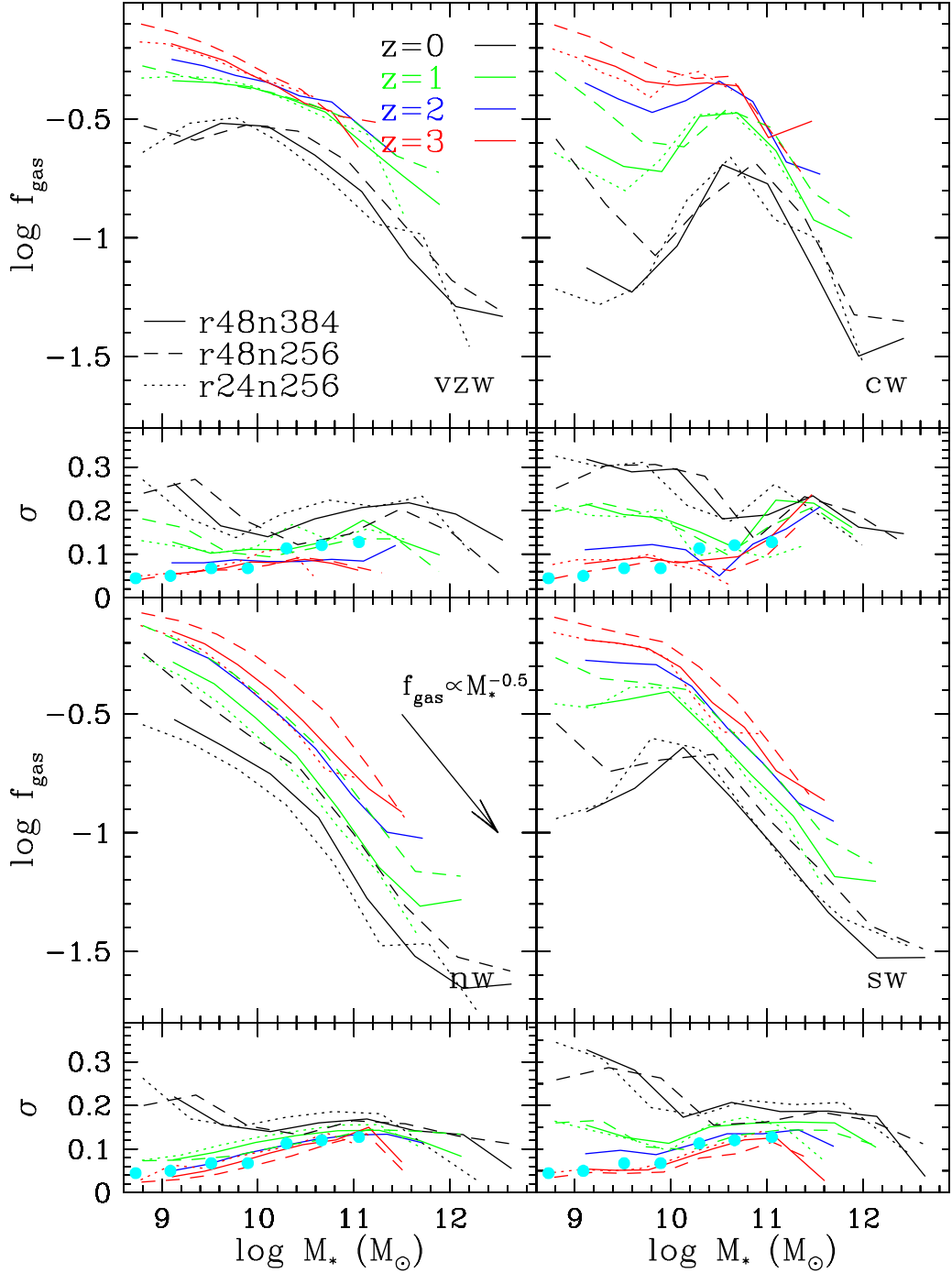}}
\vskip -0.5in
\caption{Like Figure~\ref{fig:mzevol}, only for gas fraction, showing 
the evolution of the MGR in our four wind models 
at $z=0,1,2,3$.  Cyan points show the observed scatter from \citet{pee10}.
The arrow in the no-wind panel shows a slope of $-0.5$, as is roughly
expected at the high-mass end; see text.
}
\label{fig:fgevol}
\end{figure*} 

Figure~\ref{fig:fgevol} shows the evolution of the MGR
from $z=3\rightarrow 0$ in our simulation suite.  Following
Figure~\ref{fig:mzevol}, the solid lines show medians with $1\sigma$
spread for the r48n384 runs, while the dotted and dashed lines show
likewise for the r24n256 and r48n256 runs, respectively.  Once
again, good resolution convergence is seen, as all the key features
are reproduced at each resolution, although we emphasize that the
dynamic range probed here is only a factor of eight in mass.  The
smaller panels show the running $1\sigma$ variance about the median,
and cyan points show the observed scatter from \citet{pee10}.

All models display a slowly falling gas fraction with time at a
given mass, while the variances (small lower panels) become slightly
higher with time.  Hence higher-$z$ galaxies are more gas-rich, in
qualitative agreement with observations.  The fundamental physics
governing this is a competition between gas inflow and gas consumption
rates.  In the cold accretion paradigm, the amount of gas inflowing
into the star forming region is proportional to the the gas inflowing
at the halo virial radius, since cold streams channel material to
the center of the halo~\citep{dek09}.  In detail, preventive feedback
mechanisms can reduce ISM inflow~\citep{vdv11,fau11} particularly
at low masses.

The amount of gas entering the virial radius can be estimated by the
total mass accretion rate times the baryon fraction, which scales as
$(1+z)^{2.25}$~\citep{dek09}.  Meanwhile, the gas consumption rate
is given by how fast gas can be processed into stars or an outflow,
which is given by $\tdep/(1+\eta)$.  Since $\tdep\propto t_H$, at any
given mass (which approximately corresponds to a given $\eta$), the
consumption rate is given by $t_H^{-1}$.  Since $t_H^{-1}$ evolves with
$z$ slower than $(1+z)^{2.25}$ (e.g.  $t_H^{-1}\propto (1+z)^{1.5}$
in the matter-dominated era), it is straightforward to see that the
gas supply rate drops faster with time than the gas consumption rate.
This explains why galaxies start out gas-rich but then the gas fraction
drops as the gas consumption rate catches up.

While the naive notion that galaxies simply ``consume their gas"
is qualitatively in accord with observations, the rapid consumption
times~\citep{tac10} imply that gas must be continually supplied.  In our
simulations, this indeed happens, but at a rate that becomes slower
with time (relative to the consumption rate), causing a gas fraction
that slowly drops.  The no-wind case in Figure~\ref{fig:fgevol} shows
a self-similar (in $M_*$) downwards evolution in $\fgas$ arising from
this scenario.

We can quantify these scalings and gain more insight into gas
fractions by first using the definition of $\tdep$ to rewrite
\begin{equation}\label{eqn:fgas}
\fgas = \frac{1}{1+t_{\rm SF}/\tdep},
\end{equation}
where $t_{\rm SF}\equiv M_*/$SFR=1/sSFR.  Thus the dependence of $\fgas$
on mass and redshift reflects the dependence of the ratio $t_{\rm SF}/\tdep$
on these quantities.

Let us consider the redshift evolution first.  In Paper~I we showed
that the observed $t_{\rm SF}$ at $M_*=10^{10}M_\odot$ was consistent
with following the trend predicted by halo accretion, namely $t_{\rm
SF}\propto (1+z)^{-2.25}$, from $z\sim 2\rightarrow 0$.
Figure~\ref{fig:tdep} shows that $\tdep\propto t_H$ over that time,
albeit with some deviations at low-$z$ at low masses.  Hence the
ratio $t_{\rm SF}/\tdep$ rises with time, approximately as
$(1+z)^{0.75}$ at high-$z$ and $(1+z)$ at low-$z$, resulting in a
dropping gas fraction.  Since $t_{\rm SF}$ and $\tdep$ are to rough
order independent of winds, this explains the basic behavior of
$\fgas$ dropping with time in all models.

Equation~\ref{eqn:fgas} can also be used to gain insights on the mass
dependence of $\fgas$.  At larger masses (e.g. $M_*\ga 10^{10} M_\odot$),
$\tdep\ll t_{\rm SF}$, so we can approximate $\fgas\approx \tdep$sSFR.
Consider first the no-wind case, here sSFR$\propto M_*^{-0.2}$
(approximately; see Figure~2 of Paper~I).  Combined with $\tdep\propto
M_*^{-0.3}$ (Equation~\ref{eqn:tdep}), this then roughly predicts
$\fgas\propto M_*^{-0.5}$.  This slope is indicated in the nw panel of
Figure~\ref{fig:fgevol}, and provides a good fit to the high-mass slope
of $\fgas(M_*)$.  At lower masses, once $\tdep/t_{\rm SF}$ becomes a
significant fraction of unity, the gas fraction levels off.

In the wind models, $\fgas$ will reflect features seen in the sSFR,
and at $z=0$ there are additional deviations owing to $\tdep$.  In
general, sSFRs in the wind models show a turnover to lower sSFRs
at low masses (Figure~3 of Paper~I).  The location and strength of
that turnover depends on the particular wind model, owing to wind
recycling as discussed in Paper~I.  At high masses, sSFR in wind
models is raised over the no-wind case by wind recycling, which is
is more rapid in higher mass galaxies.  At low masses, the sSFR is
lowered owing to preventive effects of winds adding energy to
surrounding gas~\citep[e.g.][]{opp10,vdv11}.  These trends are
directly reflected in $\fgas(M_*)$: At high masses, they are somewhat
larger than the no-wind case, while they all show a turnover to low
gas fractions at low masses.  At $z=0$, this reduction at low masses
is exacerbated by the drop in $\tdep$ to low masses.

An interesting regime that is not probed here is the very high-$z$ epoch
($z\ga 3$).  At sufficiently high redshifts, the rapidly-rising accretion
rate will begin to exceed the less rapidly-rising gas consumption rate.
In that case, star formation cannot keep up with the gas supply, and
the galaxies will no longer be in equilibrium.  this is then the {\it
gas accumulation phase}.  The exact epoch where this happens depends on
feedback; when outflows are highly mass loaded, the amount of infalling
gas that needs to be processed into stars is reduced, and equilibrium can
occur earlier on.  Also, there is some mass dependence because $\tdep$ has
a significant mass dependence~(Figure~\ref{fig:tdep}), but $t_{\rm SF}$
(or sSFR) is essentially independent of mass at high-$z$~\citep{gon10}.
Interestingly, there are now empirical constraints on the gas accumulation
epoch: \citet{pap11} showed, based on modeling the evolution of the
luminosity and sizes of high-redshift Lyman break galaxies, that at
$z\ga 4$ the global gas accretion rate exceeds the star formation rate,
while below that redshift they track each other.

In summary, the evolution of gas fractions reflects a competition between
cosmic inflow and gas consumption rates.  A quick estimate of their
scalings shows that cosmic inflow abates faster than gas consumption,
resulting in dropping gas fractions with time in all models.  Analytic
models based on this scenario by \citet{dut10} show a similar result.
Outflows provide higher-order modifications to this picture, particularly
owing to wind recycling and preventive feedback effects at late times.
These trends can be understood by considering the dependence of the
specific SFR (i.e. $t_{\rm SF}$) and the depletion time $\tdep$ on mass
and redshift.  At sufficiently early epochs, inflow will be so rapid
that gas processing cannot keep up, resulting in a gas accumulation phase.

\section{Comparisons to Data}\label{sec:datacomp}

\begin{figure*}
\vskip -1.8in
\setlength{\epsfxsize}{1.05\textwidth}
\centerline{\epsfbox{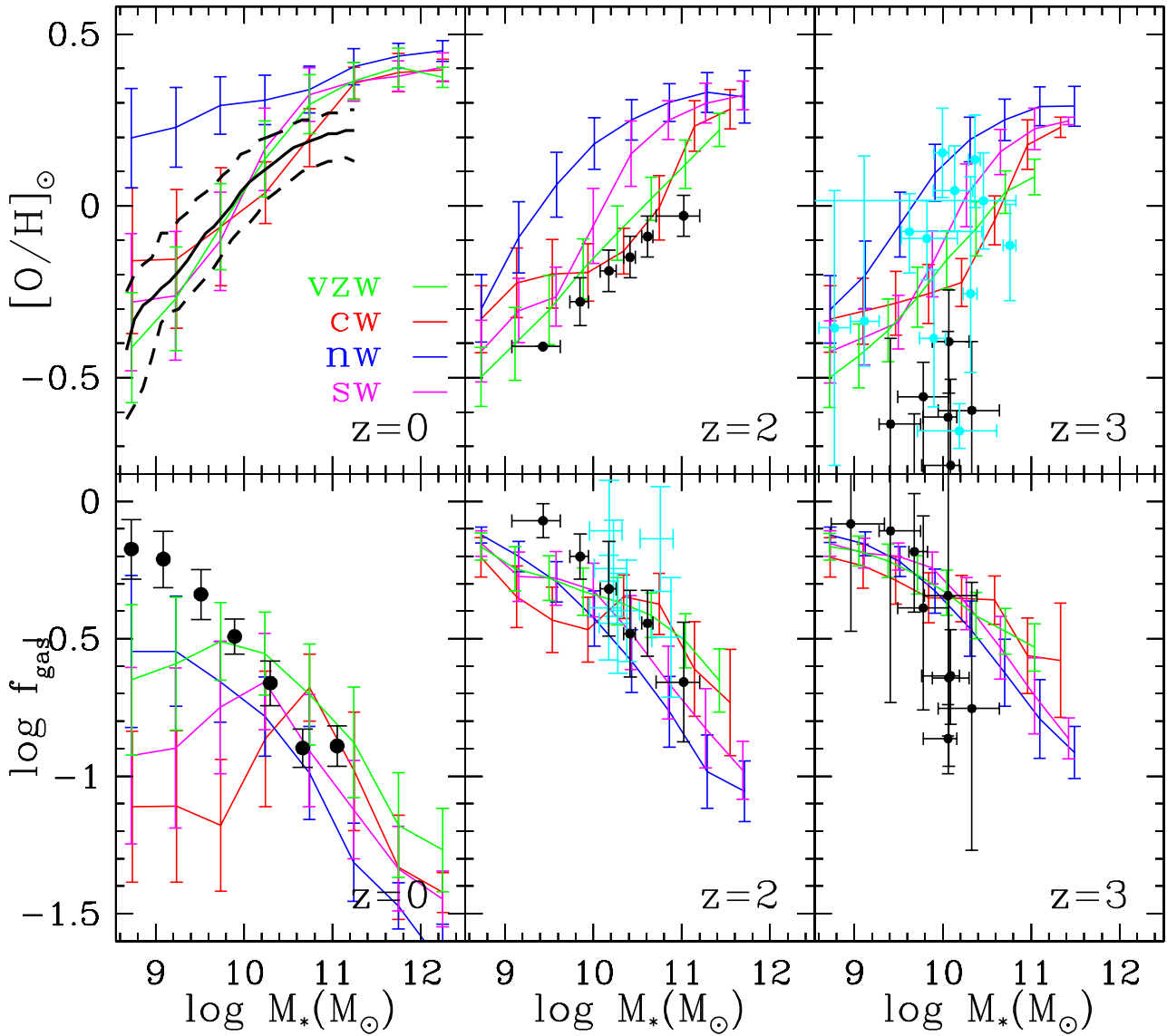}}
\vskip -1.8in
\caption{A comparison to observations for the mass-metallicity (top
panels) and mass-gas fraction (bottom panels) relation in our four
wind models at $z=0,2,3$ (left to right panels).  Lines with error
bars show the running median and $1\sigma$ variance within mass
bins from our r48n384 runs.  Data points show the MZR at $z=0$ from 
SDSS~\citep[solid
line with $16\%-84\%$ enclosing errorbars]{tre04}, at $z=2$ from \citet{erb06}
and at $z=3$ from \citet[$z\sim 3-4$; black points]{man10} and 
\citet[$z\sim 2.5-3.1$; cyan points]{ric11}.  The momentum-driven wind
scalings case comes closest to matching at both $z=0$ and $z=2$ of the
models here.  For gas fractions, $z=0$ data is
shown from a compilation by \citet{pee10}, at $z=2$, indirect gas
fractions from \citet[black points]{erb06b} and direct (CO-based)
gas fractions from \citet[cyan]{tac10}, and at $z=3$ from \citet{man10}.
The momentum-driven wind scalings provide the overall best
match, but are still discrepant at $M_*\la 10^{9.5}M_\odot$; note
that current samples measuring gas fractions generally do not include
gas-poor galaxies that are often (in our models) satellites.
}
\label{fig:Zfcomp}
\end{figure*} 

While the main purpose of this paper is to understand how galactic
inflows and outflows impact the gas and metal content of galaxies, it is
instructive to see how our suite of models compares to key observations
of these quantities out to high redshifts.  We have already seen that
different outflow models generate significantly different predictions for
gas and metal content.  Here we compare to a recent sample of forefront
observations to see how they constrain our outflow models.

Figure~\ref{fig:Zfcomp} shows a comparison of the mass-metallicity
relation (top panels) and the mass-gas fraction relation (bottom panels)
to observations (in black and cyan) at $z=0,2,3$.  The $z=0$ metallicity
observations are taken from \citet{tre04}, and the gas fraction data from
a compilation by \citet{pee10}.  At $z=2$, we show metallicites and gas
fractions from \citet{erb06}; the gas fractions here are inferred from the
star formation rate surface density and assuming the Kennicutt-Schmidt
Law.  We also show (in cyan) ``direct" gas fraction measures from
\citet{tac10} using CO measurements plus an assumed conversion of CO
to $H_2$.  At $z=3$, we compare to data from \citet{man10}, which is
from a sample of Lyman break galaxies at $z\sim 3-4$ (black points),
and from \citet{ric11} from a sample of lensed galaxies at $z=2.5-3.1$
(cyan points, metallicities only).

Looking at the $z=0$ MZR, the only model that is clearly discrepant is
the one with no winds.  Galaxies are (not surprisingly) over-enriched,
because this model greatly overproduces the stellar content of
galaxies (see e.g. Paper~I).  Other works have argued that the
$z=0$ MZR can be reproduced by varying the ISM star formation
efficiency~\citep{tas08,mou08}, or the IMF~\citep{kop07}, or
that hierarchical assembly naturally leads to this shape of the
MZR~\citep{der07}. In our simulations, the IMF is always assumed to
follow that of \citet{cha03}, and the gas depletion timescale $\tdep$
is essentially the same in all our models including no-wind.  Hence in
our models, these factors are not responsible for the differences in
the MZR.  In our equilibrium model~(eq.~\ref{eqn:mzr}), the no-wind model
overproduces metals because it does not eject ISM material to suppress
star formation.  This in some sense represents a ``cosmological" star
formation efficiency, i.e.  the amount of cosmic infall that is converted
into stars.  In our models~\citep[like those of][]{bro07}, the MZR is
governed by the cosmological star formation efficiency rather than the
ISM one.

The wind models are generally all in the ballpark of the $z=0$ SDSS
data, but they show non-trivial discrepancies.  At the massive end,
all models overproduce the metallicities in $M_*>M^\star$ galaxies.
Recall that we have arbitrarily scaled all our metallicities to
match observations at $M_*\sim 10^{10}M_\odot$, so we cannot improve
overall agreement by rescaling our yields or metallicity indicators
further.  One possible reason is that there is another feedback
mechanism that kicks in at $M_*\ga M^\star$.  As we saw in Paper~I,
this is already required for suppressing star formation in massive
systems~\citep[e.g.][]{gab11}; it would not be surprising if it
also suppressed metallicities.  We note that the observed SDSS
sample is solely emission-line galaxies, so does not include ``red
and dead" systems; nevertheless, as shown in e.g. \citet{sal07} and
Paper~I, even star-forming systems seem to have suppressed star
formation at the massive end compared with a simple extrapolation
from lower-mass systems.  Another possibility is that wind recycling
is bringing in highly enriched material that is elevating the galaxy
metallicities.  Hence this discrepancy may suggest that the impact
of wind recycling is overestimated in these simulations.  In any
case, quantitatively matching both the mass function and the MZR
plateau may provide interesting constraints on such feedback
mechanisms.

At the low-mass end, all wind models remain within the observed
$1\sigma$ variance among SDSS galaxies down to the lowest masses
probed here ($M_*\sim 10^9 M_\odot$).  However, the equilibrium
model predicts that the simulations with constant $\eta$ should
have a flat MZR at low masses (eq.~\ref{eqn:mzr}).  This flattening
trend is clearly visible in the cw run, and less so in the sw case
since it kicks in at lower masses.  Unfortunately we do not yet
have the computing power to resolve smaller systems within a
representative $z=0$ cosmological volume to confirm this prediction
directly, but higher-resolution runs done to $z=2$ do show this
trend~\citep{fin08}.  Observationally, \citet{lee06} has found that
the MZR continues with a similar slope down to $M_*\sim 10^6 M_\odot$.
This would clearly be discrepant with expectations for the
constant-$\eta$ cases, and would favor models where there is
progressively higher $\eta$ at lower masses as in momentum-driven
wind scalings.

The mass-dependent MZR evolutionary rates among wind models offer an
opportunity to constrain such models by comparisons with higher-$z$ data.
The no wind model continues to over-enrich galaxies by $\sim\times 2-3$,
but the slope is now in better agreement.  It would be suprising if
uncertianties in yields, metallicity measures, etc. could conspire to
produce a $\times 2-3$ change in metallicities, but we cannot rule out
that possibility.  But we disfavor this idea because the no-wind model
also grossly overproduces the amount of stellar mass at $z=2$ (as at
all redshifts; see e.g. Paper~I), so it would be highly surprising if
it produced the proper metal content in galaxies.

In terms of wind models, at $z=0$ the sw model was a reasonable fit, but
at $z=2$ it substantially over-enriches galaxies.  Since there are few
red and dead galaxies at this epoch, over-enrichment at the massive end
is likely to be a serious defect that will not be alleviated by quenching
(e.g. AGN) feedback.  The cw model also produces the wrong shape for
the MZR, as discussed in \citet{fin08}.  Recall that the MZR shape is
probably the most robust predictions of our simulations, so again this
is a serious defect.  Meanwhile, the momentum-driven scalings case is
a reasonable match in terms of shape and amplitude (modulo that we have
scaled all our metallicities in all wind models by a factor of 0.8).

At $z=3$, the metallicity data from \citet{man10} lies well below
all model predictions, while the data from \citet{ric11} is in
good agreement.  Since the former sample averages a slightly higher
redshift, this might imply very rapid evolution in the MZR during the
short time from $z\sim 3.5\rightarrow 2.5$, but this is very difficult
to reconcile with the observed (and predicted) slow evolution from
$z\sim 2.5\rightarrow 0$.  Models can usually match data at two of
the three epochs, but no model that we are aware of can match all
three epochs when considering the \citet{man10} data.  One possibility
is that there are observational selection effects coming into play,
because the \citet{man10} $z>3$ sample consists of rest-ultraviolet
(UV) selected galaxies, while the \citet{ric11} sample consists of
lensed galaxies.  Rest-UV selection will tend to pick out high-SFR
systems, which in turn have low metallicities for their $M_*$
(see Figure~\ref{fig:massmet}, with a more detailed discussion in
\S\ref{sec:secondpar}).  It remains to be seen if selection effects can
quantitatively reconcile the models and data.  Another possibility is that
there are differences in the poorly-understood calibrations of metallicity
indicators~\citep[e.g.][]{kew08}, although both works claim to have
carefully considered this.  Investigating the origin of the differences
in the observations is beyond the scope of this work, so we merely take
these as reflective of current systematic uncertainties in high-$z$
MZR measures.  Upcoming surveys such as the Spitzer Extragalactic Deep
Survey (SEDS) and the Cosmic Assembly Near-infrared Deep Extragalactic
Legacy Survey~\citep[CANDELS;][]{gro11,koe11}  will yield many stellar
mass-selected galaxies at these redshifts, offering opportunities through
follow-up spectroscopy to measure metallicities in large homogeneous
samples.

Turning to gas fractions, at $z=0$ all models produce the observed
trend of lower gas fractions in more massive systems.  As discussed
in \S\ref{sec:fgevol}, the trend in massive (i.e. low-$\fgas$)
galaxies is set by the depletion time times the specific SFR; these
both drop with mass, albeit mildly.  The gas fractions in the no-wind
case follow the observed shape but are too low.  One might try to
reconcile this by noting that the model definition of gas fraction
as all star-forming gas (i.e. gas above $n=0.13$~cm$^{-3}$) may not
be directly comparable to observational measures of $\fgas$.  But
at least at face value, the no-wind model appears to consume too
much of its gas into stars by $z=0$, leaving galaxies too gas-poor
and consistent with its overproduction of stars and metals.

At $z=0$, the wind models are broadly in the range of observations at
high masses, but in all cases they show a turnover in gas fractions
at low masses that is clearly in disagreement with observations.
Hence something in the current wind simulations is either removing too
much gas from dwarfs, not supplying enough gas to them, and/or consuming
their gas too quickly.  Simulations by \citet{kob07} and \citet{mou08}
likewise find that dwarf galaxies generally tend to be too old with
too little present-day star formation, so this seems to be a rather
generic problem in galaxy formation simulations within a hierarchical
paradigm.  In Paper~I we suggested that one explanation may be that the
star formation law is different in these systems~\citep[as argued by
e.g.][]{rob08}, particularly at low-$z$ when they are typically fairly
low surface brightness objects, or that the conversion of neutral into
molecular hydrogen is less efficient~\citep{kru11}.  It is curious that
the simulation without winds at least generally shows the correct trend
to the lowest masses, and hence another possibility is that preventive
feedback effects owing to winds are incorrectly modeled in these outflow
simulations.

There are other possibilities for the low-mass $\fgas$ discrepancy.
For instance, there are more quenched satellites at low masses
(Figure~4 of Paper~I) which have been depleted of gas.  However,
as we will show in \S\ref{sec:sat}, even central dwarf galaxies
show a turnover in $\fgas$.  Finally, there may be observational
selection effects in the data as gas mass observations (particularly
of small systems) tend to focus on gas-selected (e.g. \ion{H}{1})
galaxies.  For instance, to reconcile the vzw model at the lowest
mass bin ($M_*\sim 10^9M_\odot$) requires that observed galaxies
are typically $2\sigma$ outliers in gas fractions, which is not
impossible.  Upcoming surveys such as the Galex Arecibo SDSS
Survey~\citep[GASS;][]{cat10} that measure the gas content in an
unbiased sample will be helpful, but are not targetted to the lowest
mass systems where the largest discrepancies arise.  The gas content
of low-mass galaxies is therefore an important barometer for models,
and further data and modeling are needed to shed light on whether
this is indeed a serious discrepancy.

At higher redshifts, gas fraction measurements become quite uncertain.
At $z\sim 2$, all models are in fair agreement with observations,
although the data overall tend to show somewhat higher gas fractions.
Selections effects again may play a role, since these galaxies tend
to be selected as having either high gas content (so that direct
measures are feasible) or high star formation rate (which implies
high gas content; Figure~\ref{fig:fgas}).  Indirect gas fractions
as used by \citet{erb06} and \citet{man10} have additional systematic
uncertainties associated with the validity of using the Kennicutt-Schmidt
law to infer gas content, while direct gas measures must use an
uncertain conversion between CO and $H_2$ mass.  Combined with the
broadly similar $\fgas$ predictions among the models, this precludes
any meaningful constraints as of yet from high-$z$ gas fractions.
The overall rate at which gas fractions go down with time at a given
$M_*$ is consistent with observations in all models (modulo at low
masses at $z=0$), suggesting that the basic physics governing gas
fraction evolution has more to do with cosmological infall that is
same among all models as argued in \S\ref{sec:fgevol}, as opposed
to feedback mechanisms.

In summary, all simulations with outflows qualitatively reproduce
the observed evolution of metallicity and gas fraction in galaxies.
At a given $M_*$, all models produce rising metallicities and falling
gas fractions with time.  Quantitatively, the momentum-driven wind
scalings model appears to be the best overall fit to the ensemble of
observations.  Nevertheless, significant discrepancies remain even in
this case, particularly at the lowest and highest masses, which may be
owed to systematic uncertainties in metal and gas measures, observational
selection effects, and model deficiencies.

\section{Scatter \& its second parameter dependences}\label{sec:secondpar}

The small scatter in the mass-metallicity relation is a strong
constraint on galactic chemical enrichment scenarios.  While models
have historically focused on reproducing the shape of the MZR, the
reason for the small scatter, approximately 0.1~dex varying only
mildly over more than 5 dex in stellar mass~\citep{lee06}, remains
unclear.  Models invoking starburst-induced galactic outflows would
predict larger scatter in lower-mass galaxies owing to the more
stochastic nature of ejection from these systems.

As is evident from Figure~\ref{fig:mzevol}, the situation in simulations
with outflows is more complex; for instance, our momentum-driven wind
scalings model produces an MZR scatter at $z=0$ that is in good agreement
with data, and other models less so.  Here we examine what governs the
scatter in the MZR and $M_*-\fgas$ relations within the context of our
equilibrium model, particularly focusing on second-parameter correlations
of metallicity and gas fraction with star formation rate and environment.


\subsection{Star formation rate}

\begin{figure*}
\vskip -0.2in
\setlength{\epsfxsize}{0.85\textwidth}
\centerline{\epsfbox{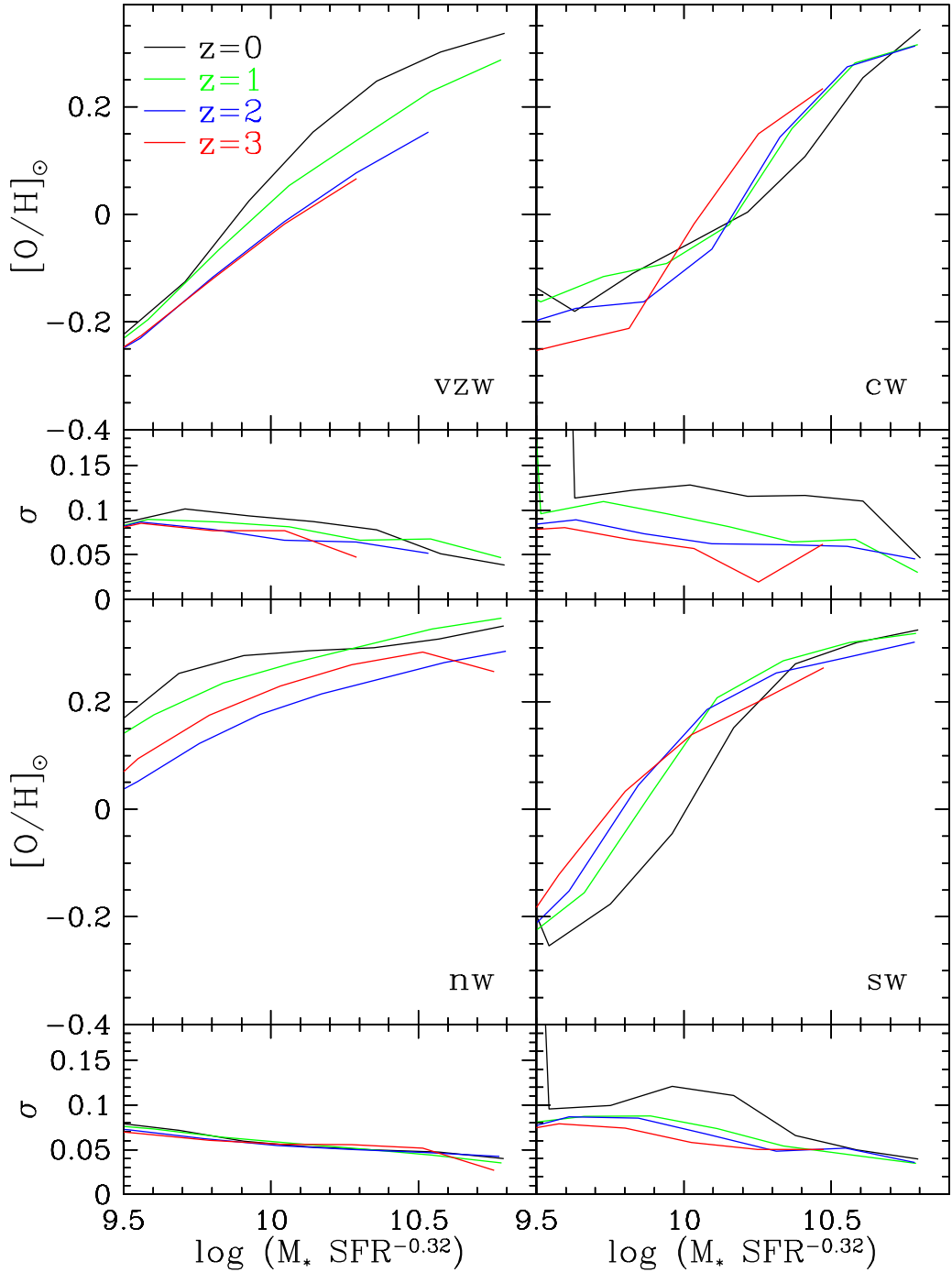}}
\vskip -0.5in
\caption{The fundamental metallicity 
relation, $Z_{\rm gas}$ vs. $M_*$SFR$^{0.32}$, in our r48n384 runs with
four wind models at $z=0,1,2,3$.  Lines with error bars show the running 
median.  Small panels below each main panel show $1\sigma$ variance.  The
variance is smaller for the FMR as compared to the MZR 
(Figure~\ref{fig:massmet}).  For comparison, observations of the FMR
by \citet{man10} found a variance of $\sigma=0.053$.
}
\label{fig:fmr}
\end{figure*} 

The mass-metallicity relation is observed to have second-parameter
dependences (i.e. correlated scatter) with other galaxy properties.
A particularly strong one that has been explored recently is the
second-parameter dependence of MZR on star formation rate.  \citet{ell08}
noticed that galaxies with higher star formation rates tend to lie below
the mean MZR.  \citet{man10} developed this idea further, and noted that a
specific combination of $M_*$ and SFR, which they called the fundamental
metallicity relation (FMR), led to a significantly smaller scatter.
\citet{lar10} independently found a similar relation.  Furthermore,
this relation appears to be invariant in redshift out to $z\sim 2.5$.
Hence the relation between mass, metallicity, and SFR provides an even
more stringent test of model predictions.

Figure~\ref{fig:massmet} shows galaxies color-coded by star formation
rate into high (blue), medium (green), and low (red) SFR relative to
the mean within each stellar mass bin.  It is a generic result from
all simulations that high SFR galaxies tend to lie below the mean MZR,
and low SFR systems lie above.  Hence the observed second-parameter
trend of the MZR on SFR is naturally reproduced in all simulations.
This suggests a fundamental process at work, independent of feedback,
that yields this trend.  We will now argue that it is a natural outcome
of the equilibrium model for the MZR discussed in \S\ref{sec:massmet}.

According to the equilibrium model, a galaxy of a given mass prefers
to be at an equilibrium metallicity that is established by its current
mass loading factor (Equation~\ref{eqn:mzr}), which sets the amount of
enrichment relative to fresh infall.  When perturbed from its equilibrium
metallicity owing to, say, a merger, this increases the galaxy mass
while lowering the metallicity (since smaller galaxies have lower
metallicities).  This moves the system below the mean MZR.  Concurrently,
it results in a more gas-rich galaxy (i.e. a high outlier in the MGR),
which in turn stimulates star formation.  This correlated behavior is
the origin of the trend that high-SFR galaxies lie at lower metallicity.

Conversely, a galaxy that suffers a lull in accretion will consume its
gas into stars and form more metals, enriching itself along the locus
of $Z\propto M_*$ (when the metallicity is significantly below the
true yield).  This is steeper than the MZR slope, and thereby the galaxy
moves above the mean MZR.  Once accretion restarts, the metals become
diluted, and the galaxy is able to return to the equilibrium relation.
In some cases, the galaxy is a small satellite dwarf that is being
environmentally quenched, and it will never restart accretion.  In the
case the system can end up quite far above the MZR, as observations by
\citet{pee08} illustrate.

From the discussion above it is clear that this scenario will yield
second parameter trends of gas fraction with SFR as well.  As seen in
Figure~\ref{fig:fgas}, galaxies with higher SFRs at a given mass tend to
have high gas fractions.  Such galaxies are undergoing an enhanced rate of
star formation relative to a typical galaxy along the $M_*-$SFR relation,
and hence will soon consume the gas and return to the mean relation.
In this way, galaxies wobble around the mean MZR, $M_*-\fgas$, and main
sequence relations owing to fluctuations in accretion.  Deviations from
equilibrium tend to return a galaxy to equilibrium, which is why this
scenario is dubbed the ``equilibrium model."  While outflows govern
the overall shape and amplitude of these relation, the qualitative
second-parameter trend with SFR is not a consequence of outflows, but
rather of equilibrium.  This is why the no-wind model shows the same
second-parameter trend as the wind models.

Outflows do, however, play a significant role in quantitatively
establishing the amount of scatter of the MZR and MGR.  A consequence of
the equilibrium model, as emphasized in~\citet{fin08}, is that the amount
of scatter reflects how fast a galaxy can return to equilibrium after
it suffers a perturbative event.  To do so, there must be sufficient
infall to re-equilibrate the galaxy.  This can be quantified by
the dilution time~\citep{fin08}, 
\begin{equation}\label{eqn:tdil}
t_{\rm dil}\equiv\frac{M_{\rm gas}}{\dot{M}_{\rm in}}=(1+\eta)^{-1}
\frac{M_{\rm gas}}{\dot{M}_*}=(1+\eta)^{-1} \tdep, 
\end{equation}
where we have used Equation~\ref{eqn:min} and the definition of $\tdep$
(eq.~\ref{eqn:tdep}).  So long as $t_{\rm dil}\la t_{\rm vir}$, where
$t_{\rm vir}$ is the dynamical time at the halo virial radius, the scatter
about the equilibrium relation will be small~\citep{fin08}.  As shown in
Figure~\ref{fig:tdep}, $t_{\rm dep}$ becomes larger at smaller masses.
If $\eta=$constant as in our cw and sw simulations, the scatter will
eventually rise significantly at low masses when the dilution time becomes
longer than the virial time.  This can be seen at the lowest masses in
the right panels (cw and sw) of Figure~\ref{fig:mzevol}.  Conversely, in
our momentum-driven wind scalings case, the trend of $\eta$ with $M_*$
mitigates the increase in $\tdep$ to low masses, and keeps the scatter
small at smaller masses.  Specifically, the vzw model (similar to all
models) yields $\tdep\propto M_*^{-0.3}$, which at $\eta\gg 1$ (small
masses) is almost exactly cancelled by $\eta\propto M_*^{-1/3}$ in this model.
So one still expects the dilution time, and hence the scatter,
to be constant to low masses.  Observations by \citet{lee06}
indicate a fairly constant scatter to low masses, though selection
effects may be artificially lowering this (H. J. Zahid, priv. comm.).
In any case, the scatter in these relations provides an independent
avenue to constrain $\eta(M_*)$.

We now examine more quantitatively the second-parameter dependence
of the MZR on SFR.  Figure~\ref{fig:fmr} shows the FMR for our
simulated galaxies, namely gas-phase oxygen abundance versus
$M_*$SFR$^{-0.32}$~\citep{man10}, with the scatter indicated in the
smaller panels below each main panel.  As can be anticipated from
Figure~\ref{fig:massmet}, the scatter is lowered using this combination
for the $x$-axis.  It is not as low as observed ($\sigma=0.053$~dex),
but a somewhat different combination of $M_*$ and SFR can lower the
model scatter further.  

Furthermore, the simulated FMR shows significantly less evolution
from $z=3\rightarrow 0$ than the MZR (note that the $y$-axis scale
is significantly smaller than in Figure~\ref{fig:mzevol}).  At
$z>3$, \citet{man10} observes the FMR to evolve strongly, which is
a consequence of the low observed metallicities in $z>3$ galaxies~(see
Figure~\ref{fig:Zfcomp}), but observations by \citet{ric11} indicate
no evolution out to $z\sim 3$.  The lack of evolution has been taken
to indicate that the FMR has a special significance that transcends
cosmic epoch.  In our models, however, the lack of (or slow) evolution
in the FMR is mostly a coincidence; it just so happens that the
increase in metallicity from high-$z$ to low-$z$ is balanced by the
evolution in typical star formation rates for that particular
combination of parameters.  The fundamental principle that drives
the FMR at any given epoch, namely the tendency for galaxies to be
drawn towards an equilibrium MZR, has little to do with the overall
evolution of the SFR at a given mass, which is set by cosmic inflow
(e.g. Paper~I).

Other observations have noted lower metallicities in merging systems.
\citet{ell08b} showed that close pairs tend to have a lower metallicity
for their mass.  It is possible that this trend is driven by the increase
in SFR in these systems as driven by the interaction, although the lack
of resolution in our models precludes us from examining this directly.
\citet{pee09} similarly found that strong outliers below the MZR tend to
be interacting galaxies and are often quite massive.  Once again this
could be related to their SFR, although massive interacting galaxies
often have significant AGN activity that produces a harder radiation
field within their ISM, which when using abundance ratios to measure
abundances can mimic a lower metallicity (C. Tremonti, priv. comm.).
While these trends are interesting, it is unclear whether they are
distinct from the overall trend of lower metallicities in higher SFR
galaxies (at a given $M_*$).  Finally, we mention that \citet{ell08}
found that systems with large half-light radii, like with high SFR, also
lie below the mean MZR.  Unfortunately, our simulations lack sufficient
resolution to robustly model galaxy sizes, so we cannot directly examine
this second-parameter dependence.

The scatter in $\fgas$ should, according to this scenario, follow the
same trend as for the metallicity.  Figure~\ref{fig:fgevol} shows that in
the compilation of \citet{pee10}, the scatter is lower at small masses.
This may be partly a selection effect, since particularly at small masses
\ion{H}{i} or CO selected samples will pick out the most gas-rich systems,
artificially lowering the scatter.  In the stellar mass-selected GASS
sample of \citet{cat10}, there is no obvious evidence for a change in
scatter in $\fgas$ versus $M_*$  among the galaxies that contain gas
(excluding gas-poor passive systems), mirroring the roughly constant
MZR scatter.  In the context of the equilibrium model, this again argues
for a mass loading factor that increases to lower masses.

In summary, galaxies in our simulations with high star formation rates
at a given mass also have lower metallicities and higher gas fractions.
This second-parameter dependence of the MZR and MGR is a natural and
straightforwardly understood consequence of the equilibrium model.
This dependence is independent of outflows, and arises purely as a
consequence of equilibrium.  Quantitatively, the dependence of scatter
on $M_*$ is a function of $\eta$ and $\tdep$, and models that have larger
mass loading factors at lower masses better match observations of constant
or mildly increasing scatter in the MZR and MGR to the lowest masses.
The FMR provides an interesting tool to quantitatively examine the
second-parameter trend with SFR, and simulations yield broadly similar
trends to those observed for the FMR.

\subsection{Environment}

\begin{figure*}
\vskip -1.0in
\setlength{\epsfxsize}{0.85\textwidth}
\centerline{\epsfbox{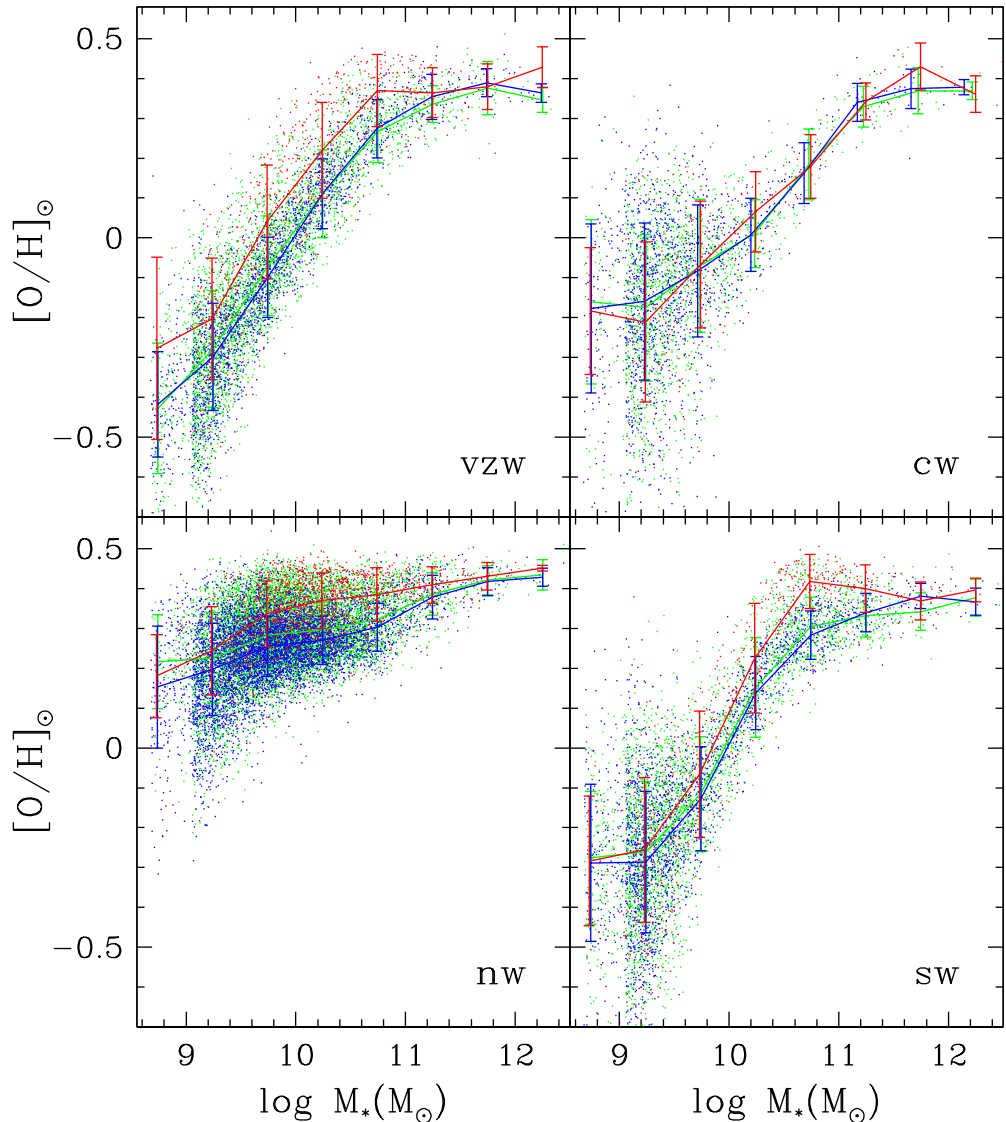}}
\vskip -1.0in
\caption{MZR at $z=0$ in our four wind models
subdivided by environment (i.e. local galaxy density in a $1\hmpc$ sphere):
High density ($>0.5\sigma$ above average; red), 
low density ($>0.5\sigma$ below average; blue), and middle (within $0.5\sigma$
of average; green).
}
\label{fig:mzenv}
\end{figure*} 

Another second-parameter dependence of the MZR was noted by \citet{coo08}
and \citet{ell09}, who showed that galaxies within dense environments
such as groups and clusters tend to have higher metallicities.
Galaxies outside such environments, in contrast, tend to have no
obvious dependence on local galaxy density.  This trend is over and
above any trend associated with star formation.  Our $48\hmpc$ volume
has some galaxy groups up to virial masses of $\sim 10^{14}M_\odot$,
but not enough to compare directly to observations of clusters.  However,
we can examine the trend in metallicity with environment as measured by
local galaxy density.

Figure~\ref{fig:mzenv} shows the $z=0$ MZR for our four wind models,
where we have subdivided galaxies by local galaxy density as measured
in a $1\hmpc$ tophat sphere.  Galaxies at densities $>0.5\sigma$
above the mean are shown in red, $<0.5\sigma$ below the mean in
blue, and those in between in green.  A running median for the MZR,
with $1\sigma$ variance, is shown for each sub-population.

The momentum-driven scalings and no-wind cases display the observed
trend: Galaxies in high-density regions lie above the mean MZR by
$\sim 0.05$~dex, while galaxies in medium and low density regions
show no discernible difference.  The constant-$\eta$ models show
significantly less dependence on environment.  The differences
disappear at the most massive end in all models.

One reason for this dependence may be that denser environments have
more enriched intergalactic gas~\citep[e.g.][]{opp06}, so accretion
onto galaxies in those environs is likely to boost the metallicity
over galaxies in a less dense region.  We believe this is indeed
the trend responsible, but how this operates is subtle, and gives
rise to distinct features among the wind models.  

The model trends are best understood if the metallicity of infalling gas
is governed by wind recycling.  Recall that the IGM is almost entirely
enriched by winds~\citep[e.g.][]{opp06}, and so any metals falling
back into galaxies constitutes wind recycling.  At high masses, wind
recycling is so effective that all galaxies re-accrete their ejected
material quickly~\citep{opp10}, which is like having no winds at all, so
the metallicity approaches the overall yield regardless of environment
(modulo an increase due to enriched infall; see \S\ref{sec:galevol}).
At sufficiently low masses, all ejected material escapes, and hence
the infalling material is mostly primordial regardless of environment.
But in the intermediate regime, environment plays a critical role in
slowing winds~\citep{opp08}, in the sense that denser regions slow winds
more and cause faster recycling.

As shown in \citet{opp10}, in the case of momentum-driven scalings
this intermediate regime occurs over a protracted range in $M_*$
since $v_w\propto v_{\rm esc}$; this protracted mass range is
reflected in this model's dependence of MZR on environment.  In the
constant-$v_w$ cases (cw and sw), there is only a small range of
masses between the fully-escaping and fully-recaptured regimes
(around the mass where $v_w\approx v_{\rm esc}$), which means that
the environmental dependence is only seen over $\sim 0.5$~dex in
mass.  Hence in these models, the high-density MZR actually shows
a peak in metallicity at the mass where recycling is most effective.
It occurs at higher $M_*$ in the cw case relative to sw since its
higher wind speed allows escape up to larger galaxies.

Empirically in our simulations, the dependence of metallicity on
environment seems to be only effective when the environment becomes
quite dense; medium and low density regions show no difference.
This is likely because only these dense environments have hot gaseous
halos~\citep{ker09a} that can significantly slow winds.  The no-wind
case also shows a dependence on environment that is obviously not
driven by wind recycling, but may be driven by tidal stripping (and
subsequent enriched infall) which is more effective in dense regions;
indeed, in \S\ref{sec:galevol} we show that the no-wind case at $z=0$
has non-negligibly enriched infall.  We leave a more detailed study
of these effects for the future.  Here we simply suggest that the
environmental dependence of wind recycling governs how local galaxy
density impacts the MZR, and our momentum-driven wind scalings model
generally reproduces the observed trend.  If this is true, studying the
environmental dependence of the MZR offers a unique probe into cycle of
baryons in and out of galaxies.

\subsection{Satellites}\label{sec:sat}

\begin{figure*}
\vskip -1.0in
\setlength{\epsfxsize}{0.85\textwidth}
\centerline{\epsfbox{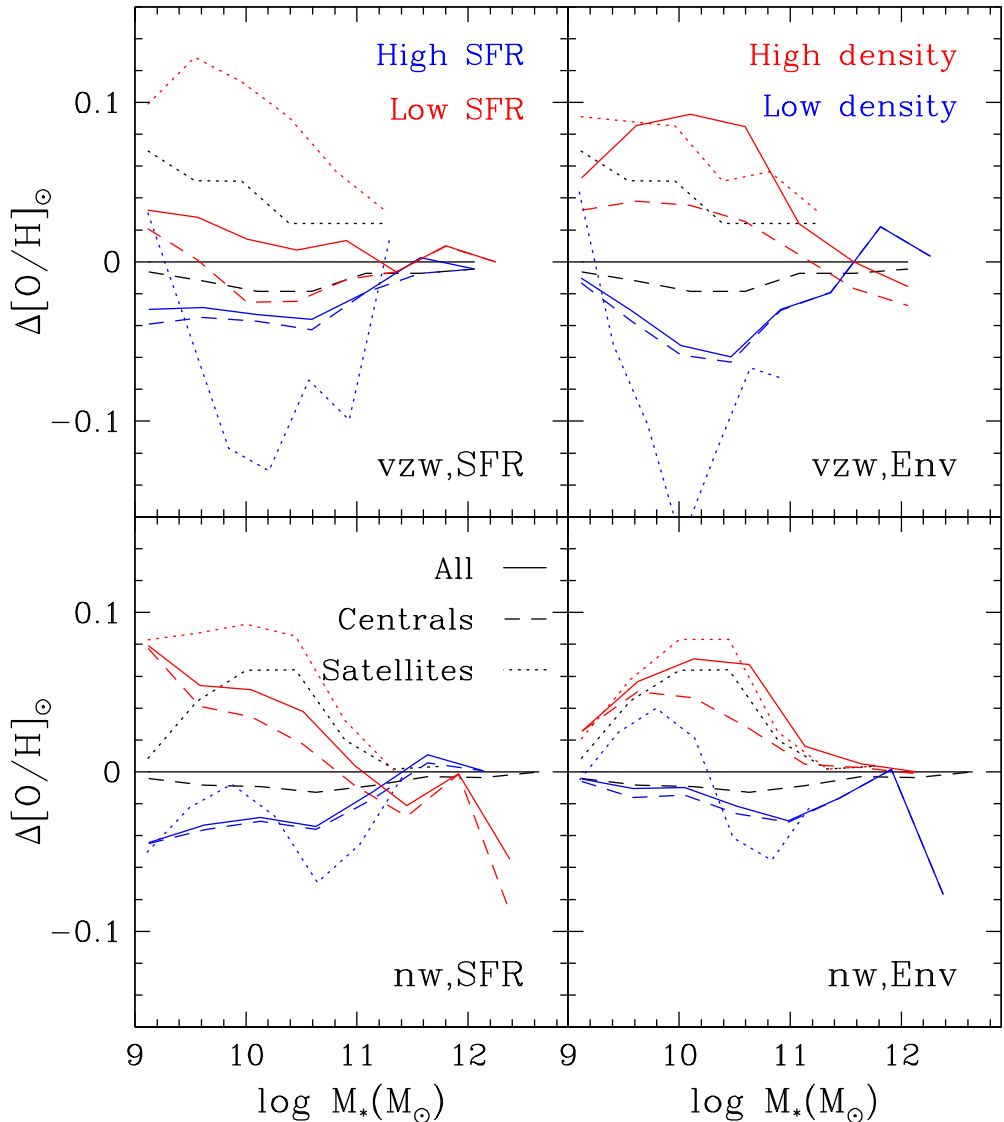}}
\vskip -1.0in
\caption{Second-parameter dependences of the MZR subdivided by
satellite vs. central galaxies at $z=0$ in our simulations.  Each curve 
shows the {\it difference}
between the MZR for a particular subsample of galaxies; the solid
black curve at zero shows the overall MZR for that model.  Top panels show the
momentum-driven wind scalings case, bottom panels the no-wind simulation.
Left panels show galaxies subdivided by star formation rate
(at a given mass) above and below the mean.  Right panels show
galaxies analogously subdivided by environment.  The blue
curves show galaxies with above-median SFRs (in left panels) and
below-median environments (right panels), red curves show below-median
SFRs and above-median environments.  Dashed and dotted black
curves show the MZR for satellites and centrals, respectively.
These populations are further subdivided into high-SFR/low-environment
(dashed and dotted blue curves) and low-SFR/high-environment 
(dashed and dotted red curves).
}
\label{fig:mzsat}
\end{figure*} 

Satellite galaxies are seen to have higher metallicities at a given
stellar mass~\citep{pee09}.  The origin of this trend is qualitatively
understood within the equilibrium model: Satellites tend to live in dense
regions where the infalling gas is more enriched, and furthermore they
are not straightforwardly fed by cold streams since they do not lie at
the bottom of the halo's potential well and hence evolve upwards off
the MZR.  In this section we quantitatively assess the differences in
satellite versus central populations for the various second-parameter
trends we have examined above.

Figure~\ref{fig:mzsat} shows the median mass-metallicity relations for
a variety of second-parameter dependences.  In order to more easily
see the dependences, we have subtracted out the overall MZR from each
subsample's MZR.  The plot shows $\Delta$[O/H]$_\odot$ dependences
on three variables:  Star formation rate, environment, and satellite
vs. central galaxies.  For the first two quantities, we subdivide the
overall sample into ``high" and ``low", which simply means above and
below the median within each mass bin.  In the left panels, we show
the median (differenced) MZR for galaxies with high-SFR as blue, and
low-SFR as red.  In the right panels, we analogously show the MZR for
high-density environment galaxies in red, and low in blue.  Top panels
show the vzw simulation, and bottom panels show the no-wind run.

We further explore the dependence of SFR and environment within satellite
(dotted lines) and central (dashed) galaxy samples.  This is done for the
full sample (black), high-SFR/low-density (blue) and low-SFR/high density
(red).  The solid line at 0 represents the original MZR of all galaxies.
In all, this figure shows how the dependences on SFR and environment
interplay with the distinction between central and satellite galaxies.

Let us begin examining Figure~\ref{fig:mzsat} by considering the blue
and red solid curves, i.e. the MZR subdivided by SFR and environment.
These trends have been noted earlier, but here the differences
are more visible since we have subtracted off the overall trend.
Comparing the solid red and black curves, we see that, as before,
low-SFR and high-environment galaxies lie above the global MZR, at least
for galaxies with $M_*\la 10^{11}M_\odot$.  The trend with environment
is generally stronger than the trend with SFR, which may be surprising
given that this second-parameter dependence has received less attention in
the literature.  In the context of the equilibrium model, this indicates
that denser environments result in significant suppression and enrichment
of inflow, such that galaxies within dense regions both have lower star
formation rates (causing higher metallicities) and are accreting higher
metallicity gas.

The source of these trends is further clarified when examining the
satellite galaxy population.  Overall, satellite galaxies (dotted
curves) show elevated metallicities at a given mass compared to
central galaxies (dashed curves), mimicking observed trends~\citep{pee09}.
Since satellites have typically lower SFRs (Paper~I) and also tend
to live in denser environments, both second-parameter dependences
discussed previously could be contributing the satellites' higher
MZR.  A particularly striking result is that the second-parameter
dependence on SFR is fairly small in the central galaxies, and is
dominated by the satellite systems.  That is, there is an enormous
difference between the metallicities of high-SFR satellites (dotted
blue) versus low-SFR satellites (dotted red) over most of the
sub-$M^\star$ mass range.  The trend is very strong in the vzw run,
but also present in the nw run.  The detailed physical origin of
this we leave for future work, but here we speculate that galaxies
first entering another halo will have enhanced SFR due to interactions
and therefore will have lower metallicities, but will then get
quenched and quickly build up metallicity to go above the mean MZR.
In any case, it is evident that in these simulations, the
second-parameter dependence of the MZR on SFR is driven more by the
satellite galaxies, while centrals show only a modest such dependence.

Looking at satellites vs. centrals subdivided by environment, it
is not as clear here what is driving the overall trend.  Both
centrals and satellites show higher metallicities in dense regions.
In the vzw model, the trend is somewhat stronger for satellites,
but not as dramatically as in the case of SFR.  Satellites are
affected by both diminished (from strangulation) and enriched inflow,
while central galaxies should not have inflow enriched, only
diminished since they should still accrete gas normally at the
bottom of the halo's potential well regardless of environment.  That
the strength of the affect is only slightly stronger for satellites
indicates that the majority of the change to the MZR in dense regions
arises because inflow is more enriched owing to residing in a denser
region.  It is worth noting that these changes in metallicity are,
in an absolute sense, not large: they are typically below 0.1~dex.
Therefore even a modest metallicity-density gradient in the
IGM~\citep{opp11} could preferentially cause galaxies in denser
environments to be over-enriched by this amount.

In summary, both environment and star formation rate play an important
role in driving second-parameter trends in the MZR.  The majority
of this trend is driven by satellite galaxies, as central galaxies
are less affected by these second parameter trends.  This shows,
as expected, that satellites have greater fluctuations in their
accretion rates that drive star formation, and are more impacted
by environmental effects.  The overall trends are consistent with
expectations from the equilibrium model, being driven by a competition
between recent accretion and outflows.  In the case of satellites
vs. centrals at a given stellar mass, the outflow rates are similar
in the simulations (being zero in the no-wind case), but the accretion
rates can vary owing to both general stochastic fluctuations and
environmental suppression.  The typical difference in metallicities
between satellites and centrals is quite small, typically of order
0.1~dex, so particular care is needed to tease out such effects in
observed samples.

\section{Evolution in Gas \& Metal Content}\label{sec:galevol}

\begin{figure}
\vskip -0.1in
\setlength{\epsfxsize}{0.5\textwidth}
\centerline{\epsfbox{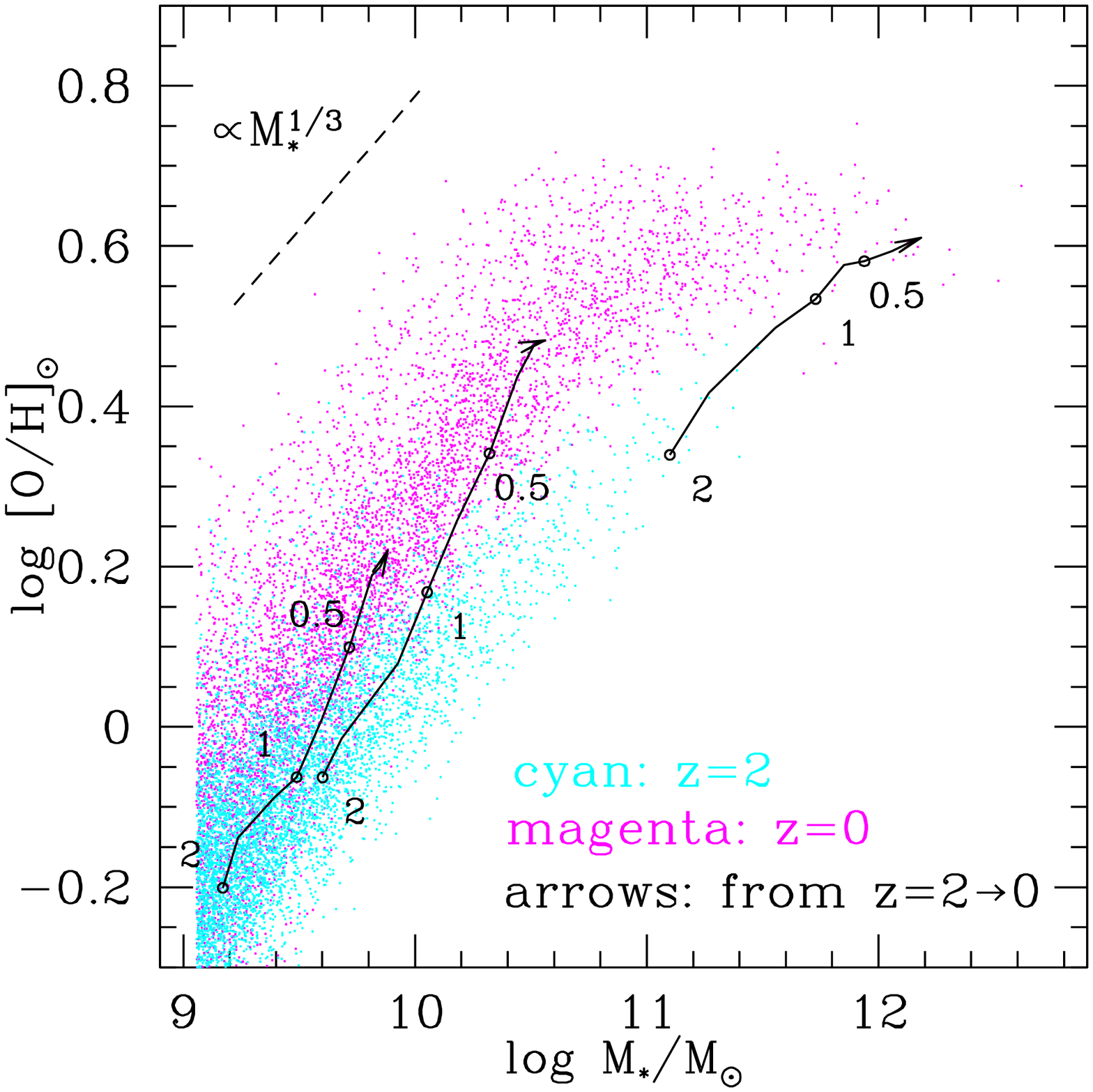}}
\setlength{\epsfxsize}{0.5\textwidth}
\centerline{\epsfbox{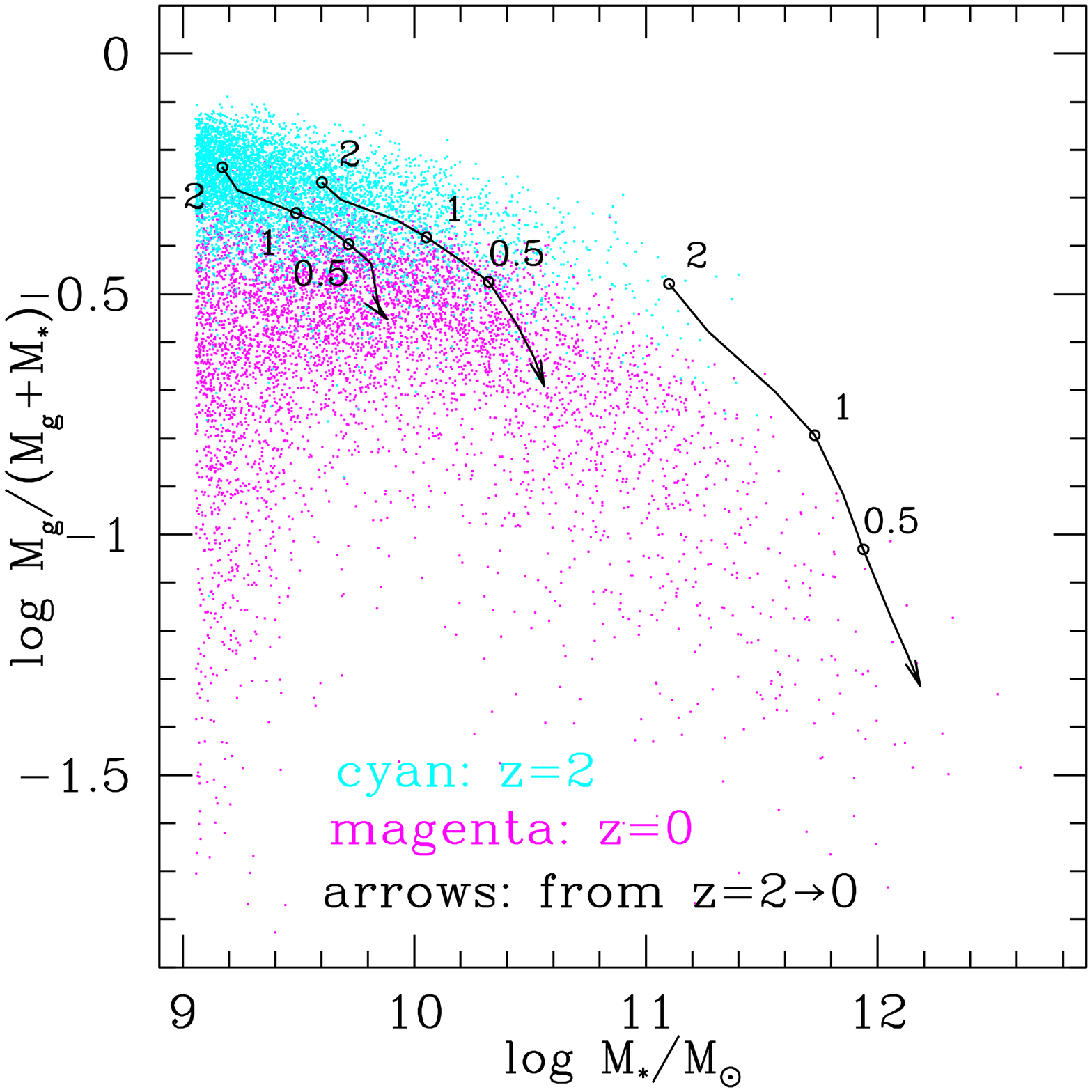}}
\caption{Evolution from $z=2\rightarrow 0$ of mean metallicity (top panel) 
and gas fraction (bottom) versus stellar mass for a set of galaxies
within 3 mass bins in our r48n384vzw run.  Cyan and magenta points show 
the overall galaxy population at $z=2,0$, respectively.  Numbers along 
the tracks indicate the redshift; tracks end at $z=0$.
}
\label{fig:metmass}
\end{figure} 

We have previously studied the evolution of the overall MZR and MGR
relations, finding that at a given mass, metallicities rise slowly
and gas fractions fall slowly with time.  In this section we examine
in more detail how particular galaxies evolve within these relations, 
in order to better understand the nature of the overall evolution.

Figure~\ref{fig:metmass} shows galaxy tracks in MZR space (top panel)
and MGR space (bottom panel) from $z=2\rightarrow 0$.  We choose
our vzw model since it does the best job of matching observations (with
notable exceptions) of the models considered.  To make these tracks,
three representative stellar masses were selected at $z=0$ (specifically
$10^{12.13}, 10^{10.57} 10^{9.85} M_\odot$), and 20 galaxies were chosen
closest to each mass.  The main progenitors of each galaxy were identified
in each output back to $z=2$, where the main progenitor is the galaxy
at an earlier epoch hosting the largest fraction of the final galaxy's
particles.  The tracks shown are the mean value of the 20 progenitors.
Numbers along the tracks indicate the redshift, with the tracks ending
in an arrow at $z=0$.

The most evident trend is that galaxies tend to evolve mostly {\it
along} the mean MZR and MGR relations, as noted by \citet{bro07}
for the MZR.  This directly translates into a slow evolution for
these relations.  The fact that galaxies move along these relations
has sometimes been forwarded as the ``cause" for the slow evolution,
but this merely begs the question, why do galaxies tend to move
along these relations?

To answer this, let us first consider the MZR.  It is straightforward
to differentiate Equation~\ref{eqn:mzr} to show that if $\eta\propto
M_*^{-x}$, then $d\log{Z}/d\log{M_*} = x$ when $\eta\gg 1$; this
is the equilibrium model prediction for the slope of the galaxy
track in MZR space at low masses, assuming that the constant of
proportionality for $\eta$ for a given galaxy is unevolving\footnote{This
turns out to be basically true, although it is something of a coincidence:
Equation~\ref{eqn:sigma} shows that for a given galaxy mass, $\sigma$
drops with time, implying a higher $\eta$ and thus a lower metallicity.
However, this is countered by the fact that a given galaxy's $\sigma$
increases with time as it grows.  It happens that the two effects
mostly cancel out for any given galaxy.}.  Hence if a galaxy obeys
Equation~\ref{eqn:mzr} at all times, it will evolve directly along
the MZR relation when $\eta\gg 1$.  For example in the vzw case,
$x=1/3$, which is identical to the MZR slope in the low-$M_*$ regime.

However, Figure~\ref{fig:metmass} shows that the evolutionary slope
is steeper than this: for the lower mass bins, $d\log{Z}/d\log{M_*}
\approx 0.6$.  The more rapid evolution must arise because an
assumption in Equation~\ref{eqn:mzr} is violated.  In particular,
it turns out that the infall is not pristine as assumed in that
equation.  \citet[Figure 3]{opp11} shows that the metallicity just
outside star-forming regions (i.e. at $n_H\approx 0.13$~cm$^{-3}$)
rises substantially from $z=2\rightarrow 0$, and exceeds solar
today.  They argue that this arises owing to the preponderance of
recycled wind accretion at later epochs, which causes the inflow
from the IGM to be increasingly enriched.

\begin{figure}
\vskip -0.4in
\setlength{\epsfxsize}{0.65\textwidth}
\centerline{\epsfbox{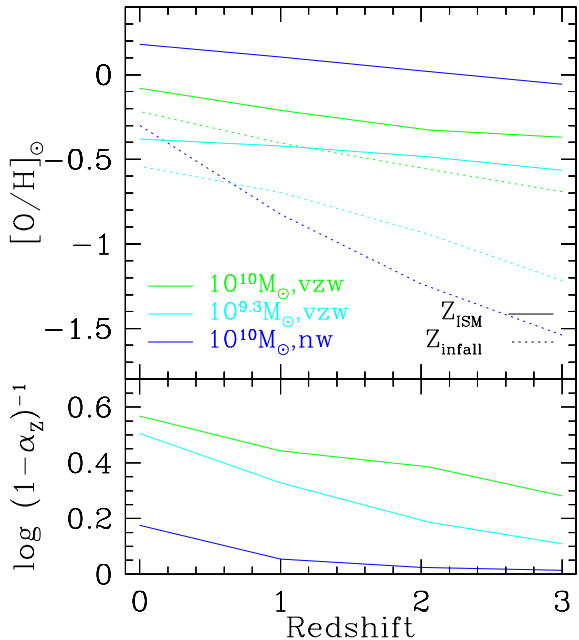}}
\vskip -2.2in
\caption{Top panel shows the evolution from $z=3\rightarrow 0$ of
metallicity in the ISM $Z_{\rm ISM}$ (solid lines) and metallicity
of infalling gas $Z_{\rm infall}$ (dotted lines).  Infalling gas is
all gas within 30 comoving kpc that is not star-forming and is moving
towards the galaxy.  Green and cyan curves show results for the vzw
simulation at stellar masses of $2\times 10^{9}$ and $10^{10} M_\odot$,
respectively.  The blue curves show the results for the no-wind case
at $10^{10} M_\odot$.  Bottom panel shows the extra multiplicative
term in the equilibrium model MZR (Equation~\ref{eqn:mzralpha}) that
accounts for enriched infall, namely
$(1-\alpha_Z)^{-1}$ where $\alpha_Z\equiv Z_{\rm infall}/Z_{\rm ISM}$.
This shows that enriched infall is
primarily responsible for the upward evolution of the MZR in our simulations.
}
\label{fig:infallmet}
\end{figure} 

Following \citet{fin08}, we can extend Equation~\ref{eqn:mzr} to
include the effects of enriched infall.  If we define $\alpha_Z$
as the ratio of infalling gas metallicity ($Z_{\rm infall}$) to the
metallicity within the ISM of the galaxy ($Z_{\rm ISM}$), then
\begin{equation}\label{eqn:mzralpha}
Z = \frac{y}{1+\eta}\frac{1}{1-\alpha_Z}.
\end{equation}
We can directly measure $\alpha_Z$ in our simulations.  For a given
mass, we take 50 galaxies near that mass and compute the mean metallicity
within the ISM (i.e. star-forming) gas.  We then compute the mean metallicity
in infalling gas.  We define infalling gas as all gas within 30~kpc (comoving)
that is {\it not} star-forming and is moving towards the galaxy (i.e. 
{\bf v}$\cdot${\bf r}~$<0$).  We also tried scaling the infall radius
with the virial radius at different masses, with only minor differences.

The evolution of $Z_{\rm ISM}$ and $Z_{\rm infall}$ are shown as the
solid and dotted lines in the top panel of Figure~\ref{fig:infallmet}.
We show the vzw model at two masses, and the no-wind case at
$M_*\approx 10^{10}M_\odot$.  The metallicity is higher around larger
galaxies, as expected.  The interesting trend is that $Z_{\rm infall}$
increases faster than $Z_{\rm ISM}$.  This is quantified in the bottom
panel where we plot $(1-\alpha_Z)^{-1}$, which is the extra factor in
Equation~\ref{eqn:mzralpha} accounting for enriched infall.  The key point
is that, from $z=2\rightarrow 0$, this term increases by $0.2-0.3$~dex.
This is identical to the excess increase in the galaxy metallicity above
simply moving up along the MZR (Figure~\ref{fig:metmass}).  The perhaps
surprising implication is that the rising metallicity at a given stellar
mass is not the result of galaxies processing more gas into stars,
but rather the result of an increasing metallicity in accreted gas.

The no-wind case also shows a similar increase in $(1-\alpha_Z)^{-1}$ of
about 0.2~dex from $z=2\rightarrow 0$.  In this case, this arises because
tidal interactions distribute metals around galaxies that can later
fall back in.  This has only become prominent since $z\sim 1$.  Overall,
the enriched infall term is much lower than in the vzw case, showing
that most of the enrichment in the infall is generated by outflows.


The gas fractions of our selected galaxies also show an evolution
generally along the relation.  The evolution does become notably
steeper at low redshifts ($z\la 0.5-1$) and at small masses, which
shows that the ``turnover" in low-$M_*$ gas fractions is a late-time
phenomenon.  These dwarf galaxies are apparently depleting their
gas reservoir too quickly, and are prevented from re-acquiring their
ejected material owing to preventive feedback processes.  We leave
a fuller examination of the interplay between such feedback processes
and gas content in dwarf galaxies for the future.

In summary, the slow evolution of the MZR and MGR scaling relations
procedurally arises from the fact that galaxies tend to evolve mostly
along these relations, with only mild deviations towards higher
metallicity and lower gas fractions with time.  Within the equilibrium
model, the higher metallicities arise because gas infall is increasingly
enriched to lower redshifts, while the lower gas fractions arise because
the depletion time becomes smaller compared to the star formation
timescale.  While these trends are qualitatively consistent with the
idea that galaxies obtain a large reservoir and slowly consume their
gas (while generating metals), our simulations suggest that the actual
physics is much more complex, driven by a balance between inflow and
outflow processes.

\section{Summary}\label{sec:summary}

In this paper and Paper~I~\citep{dav11} we have presented a study of
how the stellar, gas, and metal contents of galaxies are governed by gas
inflow and outflow processes within an evolving hierarchical Universe.
In Paper~I we investigated how galactic outflows play a key role
in modulating the stellar growth of galaxies fed primarily by cold,
filamentary accretion from the IGM.  In this paper, we have shown that
such inflow and outflow processes concurrently govern the evolution of
the metallicity and gas fraction within star-forming galaxies.

The central message of these two papers is that the evolution of the main
constituents of star-forming galaxies can be broadly understood within the
context of a cycle of inflow and outflow between galaxies and the IGM.
An idea that features prominently in our models for the evolution of the
gas and metal content is the notion of equilibrium.  Galaxies prefer to
live on specific equilibrium relations between metal, gas, and stellar
content, whose forms are set by the cosmologically-evolving inflow and
outflow rates.  The inflow rate into the ISM is tied to the accretion
rate into halos, with notable departures at low masses and late epochs
owing to preventive feedback.  Meanwhile the outflow rate appears to
be most closely tied to stellar mass, since e.g. the metallicity of a
galaxy is observed to have the tightest correlation with its stellar
mass as compared to any other individual property.  Inflow fuels star
formation, whereas outflows are the central governing agent that control
how much of the inflowing material turns into stars.  An key corollary
of this scenario is that stochastic variations in the inflow rate tend to
drive galaxies back towards the equilibrium relations, resulting in small
scatters in metallicities and gas fractions that are correlated with star
formation rate.  This equilibrium paradigm can therefore quantitatively
explain the origin of the shape, slope, and scatter of the relations
between gas, metals, and stars, as arising naturally from hierarchical
galaxy growth modulated by outflows.

With that framework in mind, we summarize the key conclusions of this
paper:
\begin{itemize}

\item Galaxy metallicities are set by a balance between inflows that
provides (relatively) pristine fuel, and outflows that reduce the
cosmological star formation efficiency by ejecting fuel.  This equilibrium
can be expressed as a function of the outflow's effective mass loading
factor $\eta$ (eq.~\ref{eqn:mzr}), meaning that the stellar mass--metallicity
relation (MZR) mostly reflects the relationship between $\eta$ and
stellar mass.

\item The evolution of the MZR in this scenario is expected to occur
mostly along the relation, as confirmed by tracking simulated galaxies.
In detail there is a slow upwards evolution in metallicity at a given
stellar mass, i.e. the MZR rises with time.  We demonstrate that this is
quantitatively understood as a result of accreted gas becoming more
enriched with time.

\item Galaxy gas fractions reflect a competition between gas accretion,
as quantified by the star formation timescale ($M_*/$SFR), and gas
consumption, as quantified by the depletion time ($M_{\rm gas}/$SFR)
(Equation~\ref{eqn:fgas}).  The depletion time in our models
is set primarily by our assumed law for star formation (based on
Kennicutt-Schmidt), while the star formation time is governed by cosmic
inflow.  Since both of these timescales are relatively insensitive to
outflows, gas fractions are (in contrast to metallicities) likewise
insensitive to outflows.

\item The stellar mass--gas fraction relation (MGR) drops slowly with
time in all models.  This arises because the gas supply rate, driven
by cosmic accretion, drops faster than the gas consumption rate, which
is tied to the galaxy's dynamical time.  Constant replenishment of ISM
gas is a ubiquitous feature of these models, as appears to be required
from observations.  Galaxies individually evolve mostly along the MGR,
but drop particularly at late epochs.  

\item Wind reaccretion plays an increasingly important role in the
evolution of the MZR and MGR at late times, particularly at $z\la 1$.
Enriched inflow directly corresponds to material that was ejected at an
earlier epoch, and alters the MZR (Equation~\ref{eqn:mzralpha}).  In the MGR,
all wind models develop a turnover at late epochs and low masses, likely
reflecting the lack of re-accretion of wind material in these systems.

\item The scatter in the MZR and MGR reflects how fast a galaxy can
return to equilibrium given a fluctuation in the accretion rate,
as quantified by the dilution time given by $\tdep/(1+\eta)$.  The
scaling of the MZR and MGR scatter with mass therefore provides an
independent constraint on $\eta(M_*)$.

\item Departures from equilibrium naturally correlate with star
formation in all models, even without winds, as it is a consequence
of equilibrium rather than feedback.  Galaxies with high SFR for
their $M_*$ are predicted to have low metallicity and high $\fgas$,
consistent with observations.  Galaxies in denser regions and
satellites are also predicted to have higher metallicities as
observed, since these systems are obtaining more enriched inflow
and/or their inflow has been curtailed because they are not residing
at the center of the halo.

\item Comparing to observations of the $z=0$ MZR, the equilibrium model
with momentum-driven wind scalings predicts an unbroken power law of
$Z\propto M_*^{1/3}$ with small scatter as broadly observed, while the
constant-$\eta$ models predict a flattening of the MZR at low masses
with a strongly increasing scatter.  All models qualitatively reproduce
the observed slow evolution of the MZR upwards from $z=3\rightarrow 0$,
but the momentum-driven wind scalings model comes closest to matching
data at both $z=2$ and $z=0$.

\item Gas fractions at $z=0$ in all wind models predict a falling
$\fgas$ with $M_*$ at high masses, and a turnover to lower $\fgas$ at
the smallest masses.  The former broadly agrees with data, while the
latter is in clear disagreement with data.  This turnover is related to
a downturn in specific SFR and upturn in age in small systems (Paper~I),
and likewise may indicate that star formation in dwarfs must be delayed
on cosmic timescales, perhaps owing to a different star formation law
or less efficient conversion of HI to H$_2$ in these systems.

\end{itemize}

Considering both metallicities and gas fractions in this paper along with
stellar masses and star formation rates from Paper I, our simulations
with momentum-driven wind scalings provide the best ensemble match
to available observations of the models considered here.  Its general
success hinges on having a higher mass loading factor in smaller systems,
as well as not having a characteristic velocity scale picked out by the
wind model that would translate into distinct features in relations such
as the MZR that are clearly not observed.  In both papers, however, we
found that the model best matches at mass scales from $M^\star$ down to
moderately large dwarfs.  At larger masses, it is clear that some form
of quenching feedback is required, which is manifested here as overly
high metallicities and a dearth of gas-free galaxies.  At small masses,
Paper I showed how small dwarfs have too low star formation rates,
which is accompanied here by too low gas fractions.  Nevertheless,
this is the first cosmological hydrodynamic simulation that provides
a reasonable match to observed present-day galaxies around $M^\star$
that contain the majority of cosmic star formation.

Taking a broader view, the problem of galaxy evolution appears to be
separable into three phases, predominantly divided by halo mass, which
could somewhat fancifully be called the birth phase, the growth phase,
and the death phase.  In the birth phase, halos are small enough that
photoionization plays a critical role in retarding accretion; these are
galaxies below the so-called filtering mass~\citep{gne00}.  In the growth
phase, galaxy growth is regulated by baryon cycling, i.e. by smooth,
filamentary accretion and ubiquitous outflows that circulate mass,
energy, and metals between galaxies and the IGM.  In the death phase,
some mechanism (often associated with feedback from the central black
hole) quenches accretion into the ISM, thereby halting star formation
and creating a passive galaxy.  Thus the three phases are not only
distinct in halo mass but also have three separate dominant feedback
mechanisms that govern galaxy growth, making galaxy evolution in each
phase physically distinct from the others.

In this paper and Paper~I we have focused on galaxies in the growth phase.
In this phase, the dark matter halo virial radius as a boundary between
the galaxy and the IGM is of secondary importance: gas flows in and out
unabated through the virial radius.  Major mergers are a sub-dominant
fueling mechanism in such systems, and are relatively unimportant in the
overall star formation history.  Central black holes may be growing within
these systems, but they play a minor role in the overall evolution.
In contrast, the formation and evolution of passive ``death phase"
galaxies appears to be critically linked to major mergers, black holes
with associated feedback, and the presence of a stable hot gaseous
halo~\citep[e.g.][]{ker05} that demarcates the virialized region from
the ambient IGM~\citep[e.g.][]{cro06,bow06,sij07,hop08,som08,dim08}.
The transition between the growth and death phases may be triggered by
a major merger, such that bulge formation and black hole growth are
linked~\citep[e.g.][]{hop08}, although it appears that maintaining a
red and dead galaxy primarily relies on the existence of a hot gaseous
halo~\citep{cro06,gab11}.  While these phases are physically distinct,
understanding the evolution of the galaxies in the death phase likely
requires well-known ``initial conditions" provided by galaxies in the
growth phase.

Many key aspects of the galaxy life cycle are far from being fully
understood.  It will be a transformative achievement in the galaxy
formation community when even the basic framework is in place, and
that work can begin towards a quantitative rather than qualitative
understanding of the main physical processes (analogous to the era
of precision cosmology).  In these two papers, we have illustrated how
simulations can be used to elucidate simple analytic relationships between
physical processes of inflow and outflow and the observable properties
of galaxies.  We have shown that cosmic dark matter-driven inflow
combined with the mass loading factors, wind recycling properties, and
preventive effects of outflows govern evolution of the basic constituents
of galaxies.  This provides a bridge between detailed studies of inflow
and outflow, the latter being much more poorly understood, and large-scale
surveys of galaxy properties across cosmic time.  In other words, given a
detailed model (e.g. from individual halo simulations) for outflow mass
loading, wind recycling, and preventive effects, we can now translate
that with good fidelity into predictions for the evolutionary properties
of galaxy populations.  Alternatively, it allows observations of galaxy
populations to be straightforwardly interpreted as constraints on the
detailed properties of inflows and outflows.  This scenario provides a
first step towards understanding the much wider range of interesting
observable galaxy properties, such as kinematics, radial gradients,
morphologies, and environmental dependences.  Continuing to grow the
synergy between multi-scale models and multi-wavelength observations
of galaxies and their surrounding gas is the best way to advance our
understanding of the life cycle of galaxies across cosmic time.

 \section*{Acknowledgements}
The authors acknowledge A. Dekel, N. Katz, D. Kere\v{s}, J. Kollmeier, C.
Papovich, J. Schaye, C. Tremonti, F. van de Voort, D. Weinberg, and
H. J. Zahid for helpful discussions, V. Springel for making \gad\ public,
and the referee for several very helpful suggestions.  The simulations
used here were run on University of Arizona's SGI cluster, ice.  This work
was supported by the National Science Foundation under grant numbers
AST-0847667 and AST-0907998.  Computing resources were obtained through
grant number DMS-0619881 from the National Science Foundation.

\end{document}